\documentclass[a4paper,oldversion]{aa}

\usepackage{graphicx,float,psfrag}
\usepackage{latexsym,amsmath,amssymb}
\usepackage{natbib}
\bibpunct{(}{)}{;}{a}{}{,}

\def \laurence  {L. Boirin}
\def \mariano {M. M{\'e}ndez}
\def \jean {J.~J.~M. In~'t Zand}
\def \laurens {L. Keek}
\def \andrew {A. Cumming}
\def \cottam {J. Cottam}
\def \lewin {W.~H.~G. Lewin}
\def \paerels {F. Paerels}

\def \strasbourg {Observatoire Astronomique de Strasbourg,
Universit\'e Louis Pasteur, 11 rue de l'Universit\'e, 67000
Strasbourg, France}

\def \sron {SRON, Netherlands Institute for Space Research,
  Sorbonnelaan 2, 3584 CA  Utrecht, the Netherlands}

\def \siu       {Astronomical Institute, Utrecht University,
                Princetonplein 5, 3584 CC Utrecht, the Netherlands}

\def \amsterdam {Astronomical Institute ``Anton Pannekoek'',
                 University of Amsterdam, Kruislaan 403, 1098 SJ Amsterdam, the Netherlands}

\def \columbia {Columbia Astrophysics Laboratory, 550 West 120th St.,
New York, NY 10027, USA}

\def \goddard {NASA Goddard Space Flight Center, Laboratory 
for X-ray Astrophysics, Greenbelt, Maryland 20771, USA}

\def \montreal {Physics Department, McGill University, 3600 Rue
University, Montreal, QC H3A 2T8, Canada}

\def \mit {Center for Space Research, Massachusetts Institute of
Technology, 77 Massachusetts Avenue, Cambridge, MA 02139, USA}

\newcounter{mycountera}
\newcounter{mycounterb}
\def \taua {$\tau_1$}
\def \taub {$\tau_2$}

\def \ta {$t_1$}

\def\countsec{\hbox{counts s$^{-1}$}}

\def \rsun {\ifmmode$R$_{\odot}\else R$_{\odot}$\fi}

\def \hcm {\hbox {\ifmmode $ atoms cm$^{-2}\else atoms cm$^{-2}$\fi}}

\def\approxgt{\mathrel{\hbox{\rlap{\lower.55ex \hbox {$\sim$}}
        \kern-.3em \raise.4ex \hbox{$>$}}}}
\def\approxlt{\mathrel{\hbox{\rlap{\lower.55ex \hbox {$\sim$}}
        \kern-.3em \raise.4ex \hbox{$<$}}}}

\newcommand {\exosat} {{EXOSAT}}

\newcommand {\xmm} {{XMM-Newton}}

\newcommand {\ie} {{\it  i.e.}}

\newcommand {\eg} {{\it e.g.}}

\def\arcsec{\hbox{$^{\prime\prime}$}}
\newcommand {\ergs} {erg~s$^{-1}$}
\newcommand {\ergcms} {erg cm$^{-2}$ s$^{-1}$}
\newcommand {\ergcm} {erg cm$^{-2}$}
\newcommand {\chisq} {$\chi ^{2}$}

\newcommand {\psame}[1] {P$_{\; \rm same}^{\; \rm {#1}}$}

\def \mxb {MXB\,1659-29}

\def \exo {EXO\,0748-676}

\def \gs {GS\,1826-24}

\def \src {\exo}

\begin{document}
 
\title{Discovery of X-ray burst triplets in \src}

\author{\laurence\inst{1} \and \laurens\inst{2,3} \and
\mariano\inst{2,4} \and \andrew\inst{5} \and \jean\inst{2,3} \and
\cottam\inst{6} \and \paerels\inst{7} \and \lewin\inst{8}}

\offprints{L. Boirin, \email{boirin@astro.u-strasbg.fr}}

\institute{\strasbourg \and \sron \and \siu \and \amsterdam \and \montreal \and \goddard \and \columbia \and \mit }

\date{Received ; Accepted:}

\authorrunning{L. Boirin et al.}

\titlerunning{Burst triplets in \src}


\abstract {Type-I X-ray bursts are thermonuclear flashes that take
place on the surface of accreting neutron stars. The wait time between
consecutive bursts is set by the time required to accumulate the fuel
needed to trigger a new burst; this is at least one hour. Sometimes
secondary bursts are observed, approximately 10~min after the main
burst. These short wait-time bursts are not yet understood.  We
observed the low-mass X-ray binary and X-ray burster \src\ with \xmm\
for 158~h, during 7 uninterrupted observations lasting up to 30~h
each. We detect 76 X-ray bursts. Most remarkably, 15 of these bursts
occur in burst triplets, with wait times of 12~min between the three
components of the triplet, T1, T2, and T3.  We also detect 14 doublets
with similar wait times between the two components of the doublet, D1
and D2. We characterize this behavior to try and obtain a better
understanding of bursts with short wait times.  We measure the burst
peak flux, fluence, wait time and time profile, and study correlations
between these parameters and with the persistent flux representing the
mass accretion rate.  (i) For all bursts with a long wait time, the
fluence is tightly correlated with the wait time, whereas burst with
short wait times generally have higher fluences than expected from
this relationship; (ii) Wait times tend to be longer after doublets
and triplets; (iii) The time profile of single bursts, S1, and of the
first burst in a double or triple burst, D1 and T1, always contains a
slow component which is generally absent in the D2, T2 and T3 bursts;
(iv) The peak flux is highest for S1, D1 and T1 bursts, but this is
still a factor of 7 lower than the highest peak flux ever seen for a
burst in this system; (v) The persistent flux, representing the mass
accretion rate onto the neutron star, is about 1\% of Eddington, which
is among the lowest value so far measured for this system.  The amount
of energy per gram of accreted mass liberated during bursts is
consistent with a fuel mixture of hydrogen-rich material. The
characteristics of the bursts indicate that possibly all bursts in
this system are hydrogen-ignited, in contrast with most other frequent
X-ray bursters in which bursts are helium-ignited, but consistent with
the low mass accretion rate in \src. Possibly the hydrogen ignition is
the determining factor for the occurrence of short wait-time bursts.
For example the 12~min wait time may be associated with a nuclear beta
decay timescale.}  
\maketitle

\section{Introduction}
\label{sec:intro}

\begin{table*}[!t]
\caption{\xmm\ observations of \src\ performed between September 19
and November 12, 2003. We indicate the observation identification and
revolution numbers, the EPIC PN start and exposure times, the number
of X-ray bursts and bursting events observed (for singlets, doublets
and triplets separately and in total), and the mean event wait time.}
\begin{center}
\begin{tabular}{l l @{\extracolsep{0.15cm}} l @{\extracolsep{0.15cm}} l @{\extracolsep{0.15cm}} l @{\extracolsep{0.25cm}} r @{\extracolsep{0.4cm}} r @{\extracolsep{0.4cm}} r @{\extracolsep{0.4cm}} r @{\extracolsep{0.4cm}} r @{\extracolsep{0.4cm}}  r @{\extracolsep{0.3cm}} r }

\hline
\hline
\noalign {\smallskip}
Obs-ID & Rev. & \multicolumn{3}{c}{Start (UT)} & $t_{\rm exp}$ & Bursts & Singlets & Doublets & Triplets & Events & Mean event wait time \\
&  & Month & Day & h:m & (h) &  & & & & & (h) \\
\hline
0160760101 & 692 &  Sept. & 19 & 13:37 & 24.6 & 10 & 6 & 2 & 0 & 8 & 3.20 \\
0160760201 & 693 &  Sept. & 21 & 13:38 & 25.1 & 14 & 6 & 1 & 2 & 9 & 2.65 \\
0160760301 & 694 &  Sept. & 23 & 10:42 & 30.0 & 14 & 7 & 2 & 1 &10 & 3.02 \\
0160760401 & 695 & Sept. & 25  & 17:29 & 20.4 & $^{a}$ 9 & 3 &$^{a}$ 3 & 0 & $^{a}$ 6 & 3.49 \\
0160760601 & 708 &  Oct. & 21 & 10:02 & 15.2 & 8 & 3 & 1 & 1 & 5 & 2.99\\
0160760801 & 710 &  Oct. & 25 & 19:19 & 17.3 & 9 & 3 & 3 & 0 & 6 & 2.71 \\
0160761301 & 719 & Nov. & 12 & 08:24 & 25.2 & 12 & 5 & 2 & 1 & 8 & 3.19 \\
\hline
& \multicolumn{4}{c}{Total} & 157.9 & 76 & 33 & 14 & 5 & 52 & Mean 3.02 \\     
\hline
\multicolumn{12}{l}{$^{a}$ including a burst detected by RGS only, a few seconds before EPIC cameras were turned on.}\\
\end{tabular}
\end{center}
\label{tab:obs}
\end{table*}

Type~I X-ray bursts are due to unstable burning of hydrogen and/or
helium on the surface on an accreting neutron star \citep[for a review
see ][]{lewin93ssr}.  Fuel accumulates for hours to days, and then
ignites as soon as the pressure and temperature conditions for
thermonuclear reactions are reached. Since the nuclear reaction rates
depend strongly on the temperature, the ignition leads to a runaway
process, and the fuel burns explosively in a rapid $\sim$10--100~s
burst releasing typically 10$^{38-39}$~ergs.  Two bursts from a given
source are usually separated by a few hours.  However, doublets with
burst intervals as short as $\sim$10~minutes have been observed from
several sources including \exo~\citep{0748:gottwald86apj},
GS~0836--429 \citep{aoki92pasj}, 4U~1608--522
\citep{1608:murakami80pasj}, 4U~1636--536
\citep{1636:ohashi82apj,1636:pedersen82apjb}, 4U~1705--440
\citep{1705:langmeier87apj} and XB~1745--248 \citep{1745:inoue84pasj}.
A burst interval as short as 50~s has been seen in \mxb\
\citep{1658:wijnands02apj}.  \cite{lewin76mnras} report SAS-3
observations of {\em three} consecutive bursts from MXB~1743-28 with
burst intervals of 18 and 4 minutes.  However, these are observations
of a field with several known bursters, and therefore there is a
slight chance that these three bursts may not be attributed to one
source.

Bursts separated by short time intervals are not well understood in
the context of the classical thermonuclear flash model.  Since the
interval is too short to accrete sufficient material to fuel a
thermonuclear burst, secondary bursts are thought to be due to burning
of residual fuel that did not burn during the primary burst. This
involves incomplete nuclear burning, fuel storage and a mixing
mechanism with the freshly accreted material
\citep{1636:fujimoto87apj}.

\exo~is one of the sources where the double burst phenomenon was best
studied thanks to several \exosat\ observations carried out at
different persistent fluxes of the system. In 1985, 26 bursts were
detected, including four doublets with a burst separation of the order
of 10--20~min \citep{0748:gottwald86apj}. The doublets only occurred
when the persistent flux was low.  As the persistent flux increased,
the doublet phenomenon stopped, the wait time of the single bursts
increased, and their shape changed from a ``slow'' (long tail) to a
``fast'' profile sometimes showing photospheric expansion.
\cite{0748:gottwald86apj} hypothesized that these variations could be
caused by the flashes changing from hydrogen-triggered hydrogen-helium
flashes at low accretion rates to helium-dominated flashes at high
accretion rates, a scenario previously theoretically
studied by \citet{fujimoto81apj} \citep[see also][]{ayasli82apj} who
outlined three ways to trigger shell flashes depending on the
accretion rate. \cite{0748:gottwald86apj} speculated that double
bursts could be a feature of hydrogen-triggered hydrogen-helium
flashes.

In another \exosat\ observation in 1986, 11 bursts were observed and
showed a regular pattern with a long recurrence time always followed
by a short one, reminiscent of the double burst phenomenon, but with
longer separations, in the range 20--70~min
\citep{0748:gottwald87apj}.  The wait time to a burst and the total
emitted energy in that burst displayed a linear relation but with an
offset energy at zero burst interval that was interpreted as
incomplete consumption of fuel in the primary burst and its subsequent
consumption in the secondary burst.  The amount of unburned fuel was
estimated to be 10--15\% of the total available nuclear energy.

 \cite{0748:cottam02nat} reported redshifted spectral lines identified
with O and Fe transitions during X-ray bursts from \src\ observed with
RGS on \xmm, implying a gravitational redshift $z=0.35$.  Combining
this redshift measurement with the peak flux of the radius expansion
burst observed by \citet{0748:wolff05apj} and the flux and color
temperature observed during the cooling tail of bursts from \exosat\
and RXTE, \citet{0748:ozel06nat} derived limits on \src\ compact
object mass and radius which ruled out several soft equations of state
for the neutron star interior.

\citet{0748:ozel06nat} further derived a lower limit to the \src\
distance of $9.2\pm1.0$~kpc (see note \ref{noteozel} in
Sect.~\ref{sec:discu}).  The source distance was previously estimated
from bursts showing photospheric expansion to range between 6.8 and
9.1~kpc \citep{jonker04mnras} or between $5.9\pm0.9$ and
$7.7\pm0.9$~kpc \citep{0748:wolff05apj}, the smallest and largest
values corresponding to hydrogen rich and poor material, respectively.
In this paper, we use the extreme values of the distance derived so
far: 5 and 10~kpc.

% dips

\src\ light curves exhibit dipping activity and eclipses every 3.8~h
 \cite[e.g. ][]{0748:parmar86apj}. Dips and eclipses are due to the
 central X-ray source being obscured by some structure above the disk,
 and occulted by the companion star, respectively, at every orbital
 period \citep[e.g. ][]{frank87aa}. Their presence indicates that
 \src\ is viewed at an inclination of $\approx 75$--$83^\circ$
 \cite[][]{0748:parmar86apj}, \ie\ almost from the accretion disk
 plane.

Here, we report the discovery of {\em triple} bursts in \src.  This is
the first time that unambiguously and repeatedly triple bursts are
detected in an accreting neutron star.  We inspect \src\ non-bursting
emission to conclude that the accretion rate is likely constant
throughout the analyzed set of \xmm\ observations
(Sect.~\ref{sec:nonbursting}). We derive the properties of the 76
bursts detected, using the 5--10~keV light curve to minimize the
contamination from the dipping activity (Sect.~\ref{sec:bursts}). We
perform a time-resolved spectral analysis of a sample of bursts not
affected by dipping.  We compare the triple bursts with the single and
double bursts detected during the same \xmm\ observations, and also
with those detected during the \exosat\ observations
(Sect~\ref{sec:exosat}). We discuss the burst properties in the
context of the well-studied helium-triggered hydrogen-helium burning
thermonuclear flash model and in the context of the poorly-studied
hydrogen-triggered hydrogen(-helium?) burning flash model
(Sect~\ref{sec:discu}).

\section{Observations and data reduction}

The \xmm\ Observatory \citep{jansen01aa} includes three
1500~cm$^2$ X-ray telescopes each with a European Photon Imaging
Camera (EPIC) at the focus.  Two of the EPIC imaging spectrometers use
MOS CCDs \citep{turner01aa} and one uses PN CCDs \citep{struder01aa}.
Reflection Grating Spectrometers \citep[RGS,][]{denherder01aa} are
located behind two of the telescopes.

\src\ was observed by \xmm\ on several occasions with different
instrument configurations.  In this paper, we use the 7 observations
performed between September and November 2003, during \xmm\
revolutions 692 to 719, for a total exposure time of 158 hours (Table
\ref{tab:obs}).  We focus on the EPIC PN data (0.1--10~keV)
that were all obtained in small window mode with the medium optical
blocking filter applied.  In this mode, only a fraction of the central
CCD chip, corresponding to 63$\times$64 pixels or
4\farcm3$\times$4\farcm4, is read out.  This allows a time resolution
of 5.7~ms to be reached, and significant photon pile-up occurs only
for count rates $\approxgt$100~\countsec.

We reduce the observations using versions 5.4.1 to 6.1.0 of the
science analysis software.  Only single and double events (patterns 0
to 4) are selected.  The ratio of background to source persistent
intensity is generally $\approxlt$1\% and reaches occasionally
$\sim$10\% during some episodes of higher background level due to
enhanced solar activity.  We do not discard any data interval for this
analysis.  For the light curves and estimate of rough burst properties
(peak count rates, number of counts in a burst), we extract source
events from within a circle of 40\arcsec\ radius centered on the
position of \src.  For the burst spectral analysis, we extract source
events from an annulus with an outer radius of $40^{\prime\prime}$ and
a $9.3^{\prime\prime}$ inner radius, in order to avoid pile-up effects
which become significant near the burst peaks. Such an annulus
contains 63\% of the events from the circular region.

\section{Non-bursting emission}
\label{sec:nonbursting}

\begin{figure}[!t]
\centerline{\includegraphics[width=0.5\textwidth]{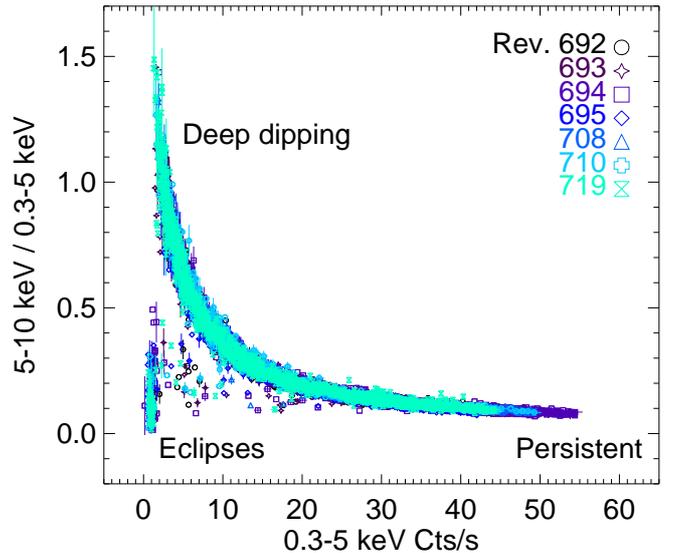}}
\caption{Color (counts in the 5--10~keV band divided by counts in the
  0.3--5~keV band) of \src\ non-bursting emission as a function of the
  soft (0.3--5~keV) intensity during the 7 \xmm\ observations whose
  revolution number is indicated in the right corner.  Error bars are
  shown on one third of the points which represent 60~s each.
   The intensity and spectral changes during revolution 719
    (bursts and eclipses excluded), and hence the boomerang track
    associated to that revolution, can be explained by changes in the
    ionization level and in the amount of absorbers in front of a
    constant underlying X-ray source \citep{diaztrigo06aa}. Remarkably,
    the data obtained during the other revolutions (spanning several
    months) follow the same track. This indicates that the spectral
    changes during these observations can be explained by the same
    phenomenon and that the underlying X-ray emission is likely the
    same during the 7 observations studied here.}
\label{fig:colorintensity}
\end{figure}

The PN light curves of the 7 observations of \src\ are shown in
Fig.~\ref{afig:rev_lc} in the appendix, in the ``hard'' 5--10~keV
(panel~a) and in the ``soft'' 0.3--5~keV energy band (panel~b), while
panel~c shows the color (counts in the hard band divided by those in
the soft band) as a function of time.  In addition to X-ray bursts,
\src\ light curves exhibit eclipses every 3.8~h and dipping activity.

Dipping is associated with spectral hardening
(Fig.~\ref{afig:rev_lc}~c), the soft light curve
(Fig.~\ref{afig:rev_lc}~b) being substantially more affected than the
hard one (Fig.~\ref{afig:rev_lc}~a). The dipping activity is highly
irregular on timescales of days, with the soft light curve shape
clearly changing from one \xmm\ observation to another.  While a
dipping pattern covering a limited phase range of $\sim$0.7--0.9 is
easily recognizable in some observations (e.g.  revolution 693), this
is not the case in others (e.g. revolution 692) where the soft
intensity and hardness display erratic variability at all orbital
phases.

\citet{diaztrigo06aa} analyzed the PN data of \src\ during
revolution 719. Excluding eclipses and bursts, they define six
intensity stages, and call the highest range the ``persistent'' level,
and the lowest one the ``Dip 5'' level. They extract one spectrum per
level and fit them simultaneously with a continuum model affected by a
neutral and by an ionized absorber. While the parameters of the
continuum are forced to be the same for the six spectra, the
parameters of the absorbers are left independent. By successfully
fitting the spectra with this method, \citet{diaztrigo06aa}
demonstrate that all the intensity and spectral changes from \src\
during revolution 719 can be simply explained by changes in the
ionization level and in the amount  of the absorbers that are
located inside the system, while the underlying X-ray source is
staying constant. Not only the ``dipping'' spectra are affected by the
local absorbers, but also the ``persistent'' one \citep[see Fig.~3 and
Table~7 of ][]{diaztrigo06aa}. The term ``persistent'', if understood
as un-affected by local absorbers, is therefore un-appropriate in this
case, \src\ being ``dipping'' (if understood as affected by local
absorbers) at all intensity levels and at all phases during that
observation.

\begin{figure*}[!t]
\centerline{\includegraphics[angle=90,width=1.\textwidth]{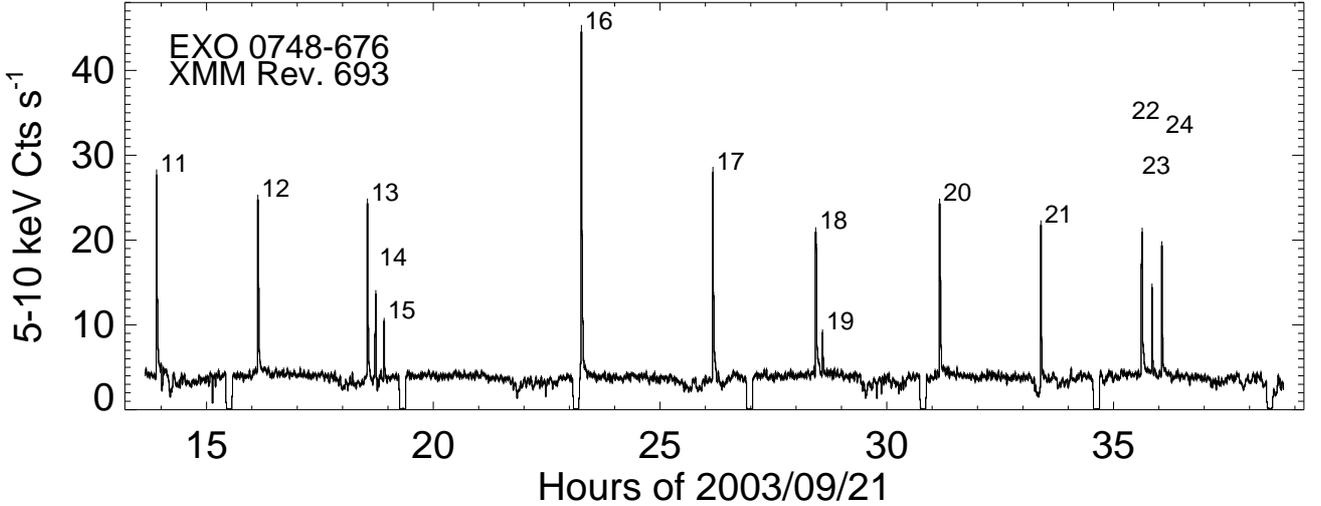}}
\caption{5--10~keV EPIC PN light curve of \src\ with a binning of 60~s during \xmm\ revolution 693. The X-ray bursts are numbered and occur either in singlets, doublets or triplets.}
\label{fig:rev_lc}
\end{figure*}

In Fig.~\ref{fig:colorintensity}, we show the
color-intensity diagram of the non-bursting emission of \src\ during
all the \xmm\ revolutions examined here (from
692 to 719). Remarkably, the data from the different observations
overlap perfectly in this diagram. Since  the intensity and
spectral changes during revolution 719, and hence the 
boomerang-like track associated with that revolution in
Fig.~\ref{fig:colorintensity}, could be explained by changing
absorbers in front of a constant X-ray source, we deduce that the
spectral changes during the other observations, and thus
their associated boomerang tracks in Fig.~\ref{fig:colorintensity},
can be explained similarly and that the underlying X-ray emission is
the same during revolutions 692 to 719. If the underlying
X-ray emission had been different from one observation to another, the
various tracks would have appeared shifted with respect to each other,
rather than being overlapping, unless a change in the underlying X-ray
emission had the very same spectral signature in the color-intensity
diagram as the change due to the absorbers, which is unlikely.

 Therefore, in this paper, we assume a constant underlying X-ray flux,
and hence a constant accretion rate, for \src\ throughout the
observations 692 to 719. We will consider the unabsorbed X-ray flux
derived by \citet{diaztrigo06aa} from revolution 719 as representative
of the unabsorbed flux for all the observations studied here.  We note
that the maximum PN count rate reached by \src\ outside bursts varies
at most by a factor~1.3 from one observation to another, which could
be considered an upper limit to the variation of the underlying flux,
if the above-mentioned interpretation of changing absorbers in front
of a constant underlying X-ray emitter throughout revolutions 692 to
719 was not correct.

Using the data from revolution 719, \citet{diaztrigo06aa} derive a
persistent 0.6--10~keV absorbed flux of $2.25 \times
10^{-10}$~\ergcms.  This corresponds to a 0.6--10~keV unabsorbed
(without any attenuation) flux of $2.81 \times 10^{-10}$~\ergcms.
Introducing a break at 50~keV \citep[as found in the 0.1--100~keV
BeppoSAX spectrum, ][]{0748:sidoli05aa} in the power-law component of
their model, we derive, in the 0.1-100~keV energy band, an
% persistent absorbed flux of $7.23 \times 10^{-10}$~\ergcms\ and an
unabsorbed flux of $8.44 \times 10^{-10}$~\ergcms\ that we consider as
the representative bolometric underlying flux of \src\ for revolutions
692 to 719.  At 5~kpc, this implies an unabsorbed luminosity of $8.44
\times 10^{35}$~\ergs\ in the 0.6--10~keV band, and $2.52 \times
10^{36}$~\ergs\ in the 0.1--100~keV band. At 10~kpc, this implies a
luminosity of $3.36 \times 10^{36}$~\ergs\ in the 0.6--10~keV band and
$1.01 \times 10^{37}$~\ergs\ in the 0.1--100~keV band.  We estimate
the relative error on the fluxes and luminosities due to the spectral
fit uncertainties to be $\sim$3\%.

\section{Bursts}
\label{sec:bursts}

76 X-ray bursts are recorded during the 7 \xmm\ observations of
\src\ (Table~\ref{tab:obs}). As illustrated in Figs.~\ref{fig:rev_lc}
and \ref{fig:burst_lc} (see also the complete set of bursts in
Fig.~\ref{afig:rev_lc} of the appendix), most bursts occur after a
wait time since a previous burst of typically 3~h, while others occur
after a wait time of only $\sim$12~min.  We will see in
Sect.~\ref{sec:paramdistr} that there is a clear distinction between
these long and short wait-time bursts.  We call several consecutive
bursts separated by short wait times a {\em bursting event}.  The
observations contain events of one, two and three bursts, which we
refer to as {\em singlets}, {\em doublets} and {\em triplets},
respectively.  The burst types are denoted by S1 for a singlet, D1 and
D2 for the first and second burst in a doublet, respectively, and T1,
T2 and T3 for the first, second and third burst in a triplet,
respectively.

 The observations contain 52 bursting events: 33 singlets, 14 doublets
and 5 triplets.  On average, one event occurs every 3~h, one singlet
every 5~h, one doublet every 11~h and one triplet every 20~h.  This is
the first time that triplets are unambiguously and repeatedly detected
in an accreting neutron star. The bursts from all the triplets are
shown in Fig.~\ref{fig:burst_profiles} together with the bursts from
one singlet and from one doublet.

To reliably compare all the bursts despite the fact that many of them
occur during dipping (Fig.~\ref{afig:rev_lc}), unless otherwise stated,
we base the bulk of our analysis on the hard 5--10~keV energy band
because it is much less contaminated by the dipping activity than the
soft band (cf. Sect.~\ref{sec:nonbursting}, \citealt{diaztrigo06aa}).
In addition, in Sect.~\ref{sec:spectral}, we perform a spectral
analysis in the full 0.1--10~keV band of a sample of bursts that are
the least affected by dipping, from which we derive conversion factors
that we use to convert the 5--10~keV parameters values into bolometric
ones in the rest of the analysis.  Bursts 26 and 38 contain data gaps
and are excluded from part of the analysis.

\subsection{Parameters definition}
\label{sec:definitions}

\begin{figure}[!t]
\centerline{\includegraphics[angle=0,width=0.46\textwidth]{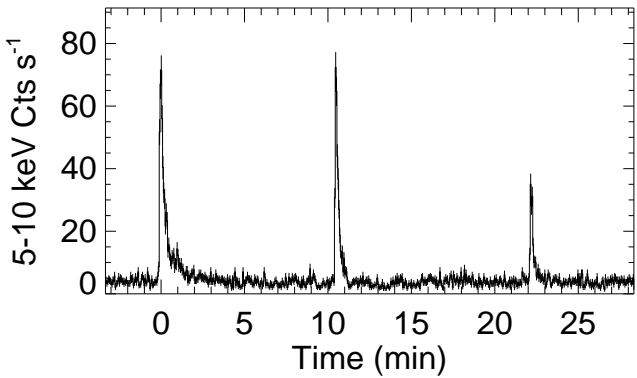}}
\caption{5--10~keV EPIC PN light curve of \src\ with a binning of 3~s
showing the triplet of bursts 13, 14 and 15.  Time is given in minutes
from the peak time of the burst 13.}
\label{fig:burst_lc}
\end{figure}

\begin{figure}[!t]
\centerline{\includegraphics[angle=0,width=0.48\textwidth]{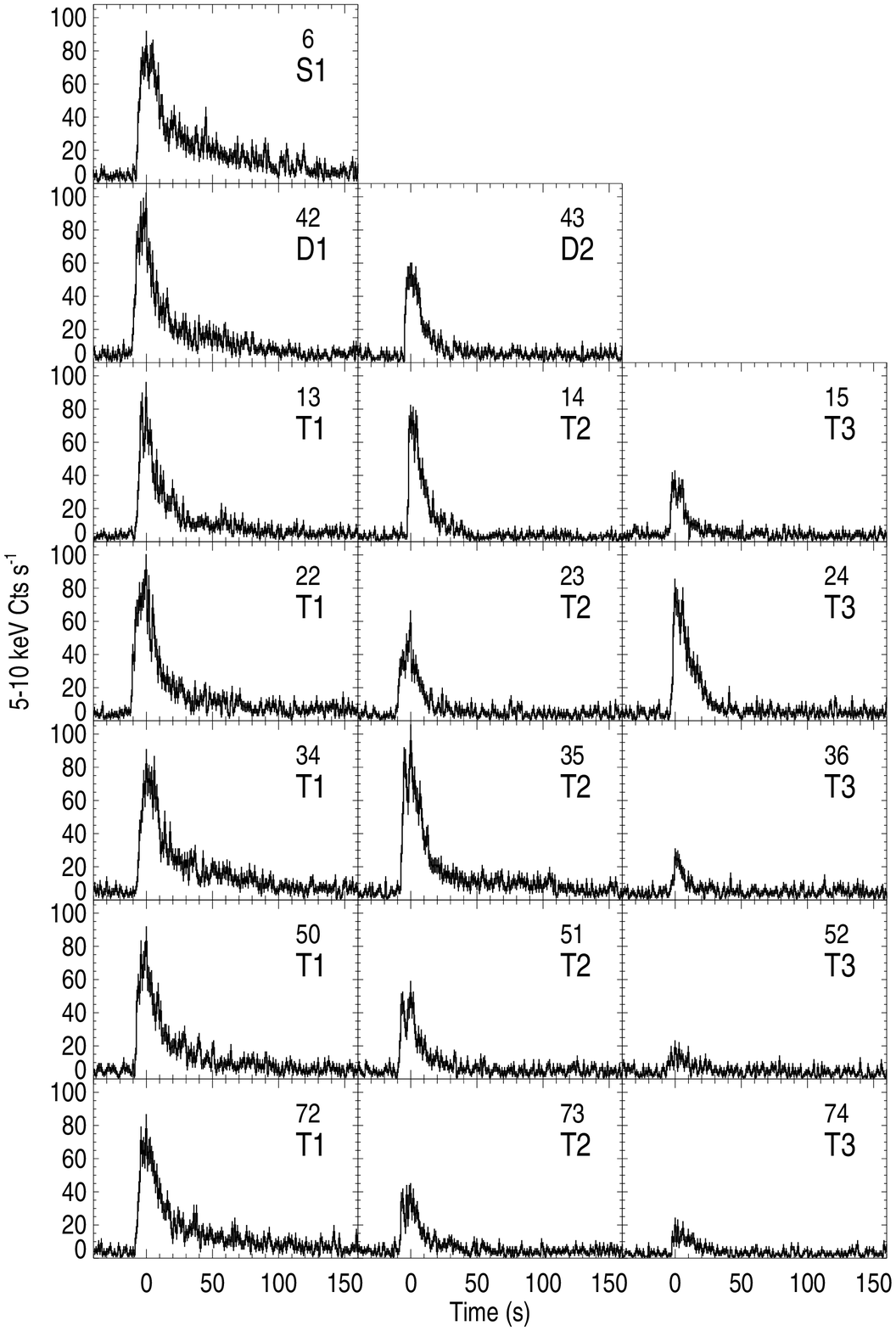}}
\caption{5--10 keV EPIC PN light curves of \src\ with a binning of 1~s
  showing the bursts profiles in one singlet, in one doublet and in
  all of the five detected triplets.  The burst number and its type
  are indicated in each panel. Time is given in seconds from the burst
  peak time.}
\label{fig:burst_profiles}
\end{figure}

Prior to each burst, we select a reference interval free of eclipse,
instrumental data gaps or other bursts, in the 1-s resolution light
curve.  We define the burst start time when the count rate reaches
3~$\sigma$ above the level in the reference segment, as the burst
rises, and the burst stop time when the intensity drops back to a
level of $1.1$ times the mean intensity in the reference segment, as
the burst decays. We define the burst duration as the separation
between these burst stop and start times.  The peak time is the time
at which the count rate is maximum, again in the 1-s resolution
5--10~keV light curve.  We define {\it the wait time} of a burst as
the separation between its peak time and the peak time of the previous
burst. We take {\it the event wait time to the next event} as the
separation between the peak time of the first burst in the event to
the peak time of the first burst in the following event.

We determine the total net counts in a burst by summing all the counts
detected between the burst start and stop times and subtracting from
this value an estimate of the number of persistent-emission counts
that were emitted over the burst duration (average persistent count
rate in the reference segment multiplied by the burst duration).  The
net counts in a burst is a tracer of the burst fluence. We define the
net counts in the event (tracer of the event fluence) as the sum of
the net counts of each burst in the event.

\renewcommand{\baselinestretch}{1}
\begin{table*}[!t]
\caption{Mean, minimum and maximum values of the burst peak flux,
duration, fluence and $\alpha$ value, for each burst type, derived
from the 1~s resolution EPIC PN 5--10~keV light curve, assuming a
conversion factor of $8.1 \times 10^{-11}$~erg~cm$^{-2}$~count$^{-1}$
between the 5--10~keV net number of counts and the bolometric burst
fluence, a conversion factor of $8.2 \times
10^{-11}$~erg~cm$^{-2}$~count$^{-1}$ between the 5--10~keV peak count
rate and the bolometric burst peak flux (Sect.~\ref{sec:spectral}) and
a bolometric persistent flux of $8.44 \times
10^{-10}$~\ergcms\ (Sect. ~\ref{sec:nonbursting}).}

\begin{tabular}{l   @{\extracolsep{0.25cm}}l@{\extracolsep{0.15cm}}l@{\extracolsep{0.15cm}}l     @{\extracolsep{0.25cm}}l@{\extracolsep{0.22cm}}l@{\extracolsep{0.22cm}}l     @{\extracolsep{0.25cm}}l@{\extracolsep{0.15cm}}l@{\extracolsep{0.15cm}}l   @{\extracolsep{0.25cm}}l@{\extracolsep{0.15cm}}l@{\extracolsep{0.15cm}}l @{\extracolsep{0.25cm}}l@{\extracolsep{0.15cm}}l@{\extracolsep{0.15cm}}l}
\hline
\hline
\noalign {\smallskip}
Type & \multicolumn{3}{c}{$t_{\rm wait}$ (h or min)} &  \multicolumn{3}{c}{Peak flux} &  \multicolumn{3}{c}{Duration (s)} &  \multicolumn{3}{c}{Fluence} &  \multicolumn{3}{c}{$\alpha$} \\
 & \multicolumn{2}{c}{} &  \multicolumn{5}{c}{($10^{-9}$~\ergcms)} &  \multicolumn{2}{c}{} &  \multicolumn{3}{c}{($10^{-7}$~\ergcm)} &  \multicolumn{3}{c}{} \\
     & Mean & Min & Max &   Mean & Min & Max &   Mean & Min & Max &   Mean & Min & Max &  Mean & Min & Max \\
\noalign {\smallskip}
\hline 
\noalign {\smallskip}
S1  &  3.20~h  &  2.23~h  &  4.64~h  & 7.2 & 5.5 & 11 & 116 & 55 & 177 & 1.9 & 0.90 & 3.1 & 53 & 42 & 102 \\ 
D1  &  2.87~h  &  2.04~h  &  3.70~h  & 7.0 & 6.1 & 8.6 & 100 & 65 & 156 & 1.7 & 1.1 & 2.2 & 52 & 45 & 61 \\ 
T1  &  2.46~h  &  2.23~h  &  2.79~h  & 6.9 & 6.4 & 7.5 & 81 & 61 & 94 & 1.4 & 1.2 & 1.6 & 54 & 49 & 57 \\ 
\noalign {\smallskip}
S1, D1, T1  &  3.02~h  &  2.04~h  &  4.64~h  & 7.2 & 5.5 & 11 & 109 & 55 & 177 & 1.8 & 0.90 & 3.1 & 53 & 42 & 102 \\ 
\noalign {\smallskip}
\noalign {\smallskip}
D2  &  13.20~min  &  8.40~min  &  19.02~min  & 4.7 & 1.7 & 7.5 & 39 & 22 & 53 & 0.61 & 0.26 & 1.1 & 12 & 6 & 17 \\ 
T2  &  12.18~min  &  9.42~min  &  15.36~min  & 5.2 & 3.2 & 7.9 & 50 & 28 & 116 & 0.73 & 0.38 & 1.5 & 10 & 5 & 16 \\ 
\noalign {\smallskip}
T3  &  11.70~min  &  10.62~min  &  13.02~min  & 2.9 & 1.6 & 6.3 & 23 & 15 & 42 & 0.30 & 0.091 & 0.87 & 35 & 8 & 63 \\ 
\noalign {\smallskip}
D2, T2, T3  &  12.66~min  &  8.40~min  &  19.02~min  & 4.5 & 1.6 & 7.9 & 38 & 15 & 116 & 0.57 & 0.091 & 1.5 & 16 & 5 & 63 \\ 
\noalign {\smallskip}
\noalign {\smallskip}
All types  &  2.05~h  &  8.40~min  &  4.64~h  & 6.3 & 1.6 & 11 & 86 & 15 & 177 & 1.4 & 0.091 & 3.1 & 40 & 5 & 102 \\ 
\noalign {\smallskip}
\hline
\noalign {\smallskip}
\end{tabular}
\label{tab:mean}
\end{table*}

We apply a multiplicative factor of $8.1 \times
10^{-11}$~erg~cm$^{-2}$~count$^{-1}$ to convert the 5--10~keV net
counts into the bolometric fluence, and a factor of $8.2 \times
10^{-11}$~erg~cm$^{-2}$~count$^{-1}$ to convert the 5--10~keV peak
count rate into the bolometric peak flux (see details in
Sect.~\ref{sec:spectral}).

The $\alpha$ value of a burst is defined as $f_{\rm p} \times t_{\rm
wait} / E_{\rm b}$, where $f_{\rm p}$ is the bolometric persistent
flux, $t_{\rm wait}$ the burst wait time and $E_{\rm b}$ the
bolometric burst fluence.  As the persistent flux depends on the
accretion rate, the product $f_{\rm p} \times t_{\rm wait} $ is a
tracer of the mass accumulated on the neutron star via accretion
between the preceding burst and the considered one. In the context of
thermonuclear flash models, the burst fluence is expected to be
proportional to the amount of nuclear fuel available. At accretion
rates where the fuel does not burn stably outside bursts, the burst
fluence is thus expected to be proportional to the accumulated mass,
and $\alpha$ to be constant, in a given bursting thermonuclear regime.
Here, we determine the $\alpha$ values assuming a constant bolometric
persistent flux of $8.44 \times 10^{-10}$~\ergcms\
(Sect.~\ref{sec:nonbursting}).

\subsection{Parameters distributions \label{sec:paramdistr}}

The peak time, wait time and type of all the bursts are listed in
Table~\ref{atab:times}. The mean, minimum and maximum values of the
burst wait time, peak flux, duration, fluence and $\alpha$ are given
in Table~\ref{tab:mean} for each burst type and for some combinations
of types.

Where relevant, we indicate the skewness and kurtosis of the samples.
The skewness of a sample $Y_1, Y_2,\ldots, Y_N$ is defined as
$\sum_{i=1}^{N}(Y_i-\bar{Y})^3/(N-1)s^3$, where $\bar{Y}$ is the mean,
$s$ the standard deviation, and $N$ the number of data points.  The
skewness is expected to be 0 for a symmetric distribution, $<0$ for a
skewed left one, and $>0$ for a skewed right one.  The kurtosis,
defined as $[\sum_{i=1}^{N}(Y_i-\bar{Y})^4/(N-1)s^4]-3$, is expected
to be 0 for a normal distribution, $<0$ for a flatter one, and $>0$ for
a more peaked one.

We use the two-sample Kolmogorov-Smirnov test to estimate the
probability, \psame{a-b}, that two samples \texttt{a} and \texttt{b}
come from the same distribution.  Since there are only 5 triplets (as
compared to 14 doublets and 33 singlets), the probabilities involving
the triplets are inferred from particularly small numbers and should
be considered with care.

\subsubsection{Wait time}

\begin{figure}[!t]
\vspace{-0.5cm}
%\centerline{\includegraphics[angle=0,width=0.45\textwidth]{fig_wait_combined.eps}}
\centerline{\includegraphics[angle=0,width=0.45\textwidth]{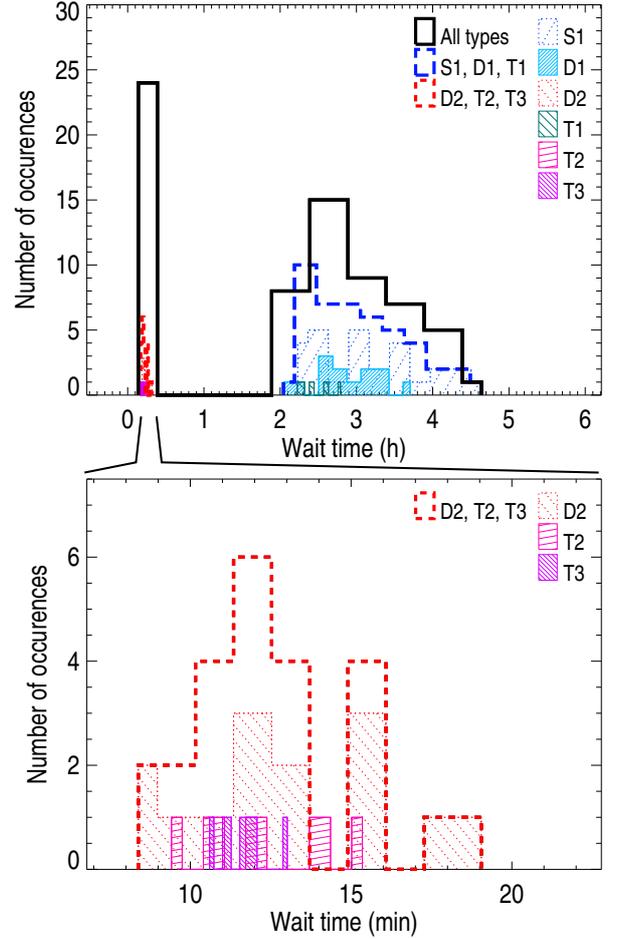}}
\caption{Histograms of the wait times. The bottom plot zooms in on the
 top one.  The histograms are built independently for each burst type
 and for some combinations of types, using always 10
 bins. Consequently, the bin size is different from one histogram to
 another. Furthermore, the size of the first and last bins is adjusted
 to match the exact observed range of wait times for a given
 histogram.}
\label{fig:wait}
\end{figure}

The histograms of the wait times are shown in Fig.~\ref{fig:wait} for
each burst type and for some combinations of burst types. The wait
times range between 8.4~min and 4.6~h, but we observe no burst with a
wait time between 20~min and 2~h. Two distinct groups of bursts are
thus clearly visible: one group made up of the S1, D1 and T1 bursts
which occur after a mean wait time of 3.0~h, and another group
consisting of the D2, T2 and T3 bursts which occur after a mean wait
time of only 12.7~min. We will refer to these two groups as the bursts
with long wait times (LWT) and short wait times (SWT),
respectively. This distinction is confirmed by the very low
probability (\psame{long-short}~=~$7 \times 10^{-14}$) that the wait
times from the two groups come from the same distribution.  The SWT
histogram and the LWT histogram have the same skewness of 0.64 , and a
similar kurtosis of -0.57 for the LWT and -0.31 for the SWT.

\begin{figure}[!h]
\vspace{-0.5cm}
%\centerline{\includegraphics[angle=0,width=0.45\textwidth]{fig_histo_waitnext_group_8bins.ps}}
\centerline{\includegraphics[angle=0,width=0.45\textwidth]{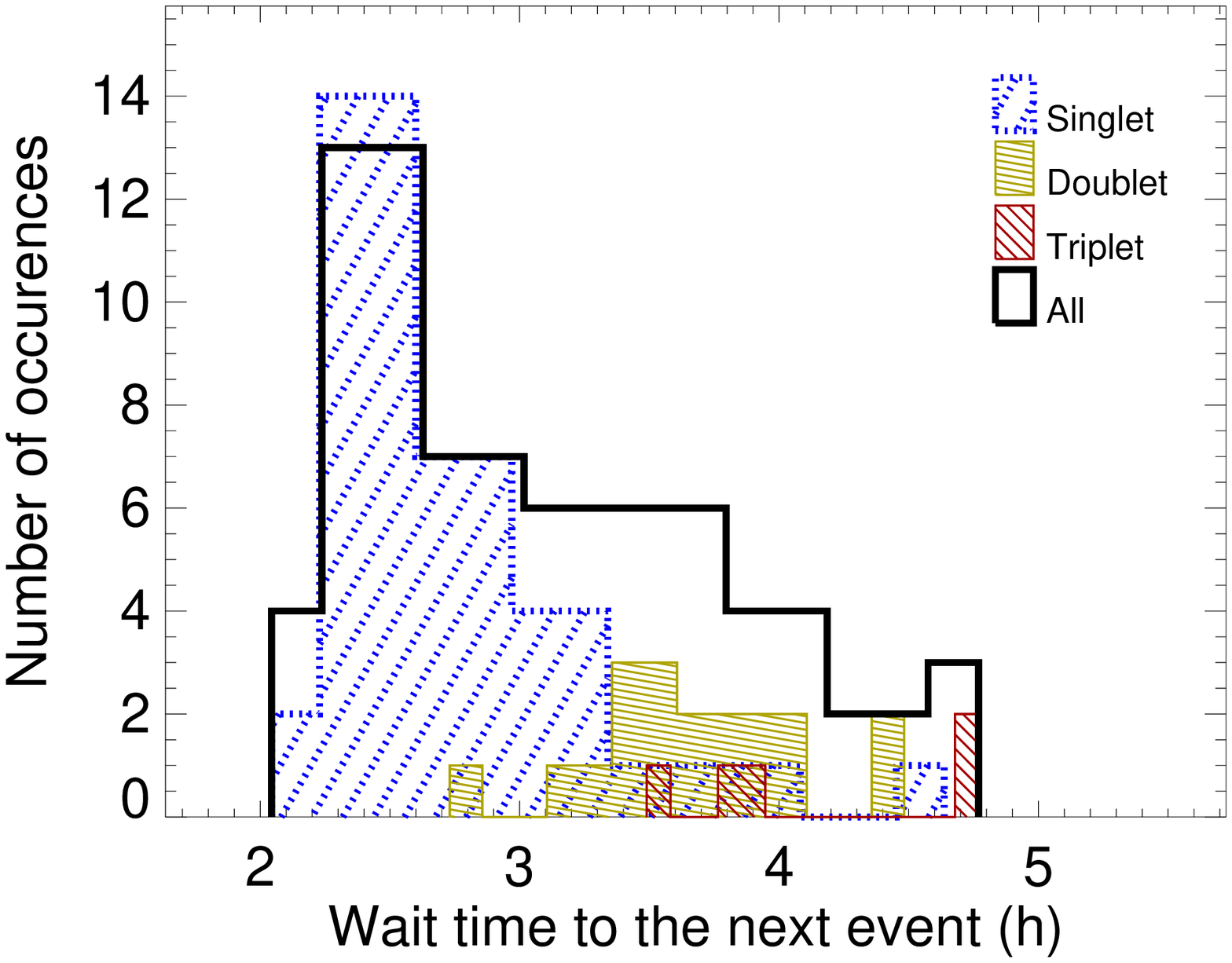}}
\caption{Histograms of the event wait times to the next event, built
independently for the singlets, doublets, triplets and for all events,
using 8 bins.}
\label{fig:waitnextevent}
\end{figure}

Inside the LWT group, the wait time distributions from two different
burst types are not significantly different (\psame{S1-D1}~=~37\%,
\psame{S1-T1}~=~23\%, \psame{D1-T1}~=~25\%).  Inside the SWT group,
the D2, T2 and T3 bursts also seem to follow the same wait times
distribution (\psame{D2-T2}~=~96\%, \psame{T2-T3}~=~70\%,
\psame{D2-T3}~=~35\%).

\begin{figure}[!t]
%\centerline{\includegraphics[angle=0,width=0.45\textwidth]{fig_histo_peak.ps}}
\centerline{\includegraphics[angle=0,width=0.45\textwidth]{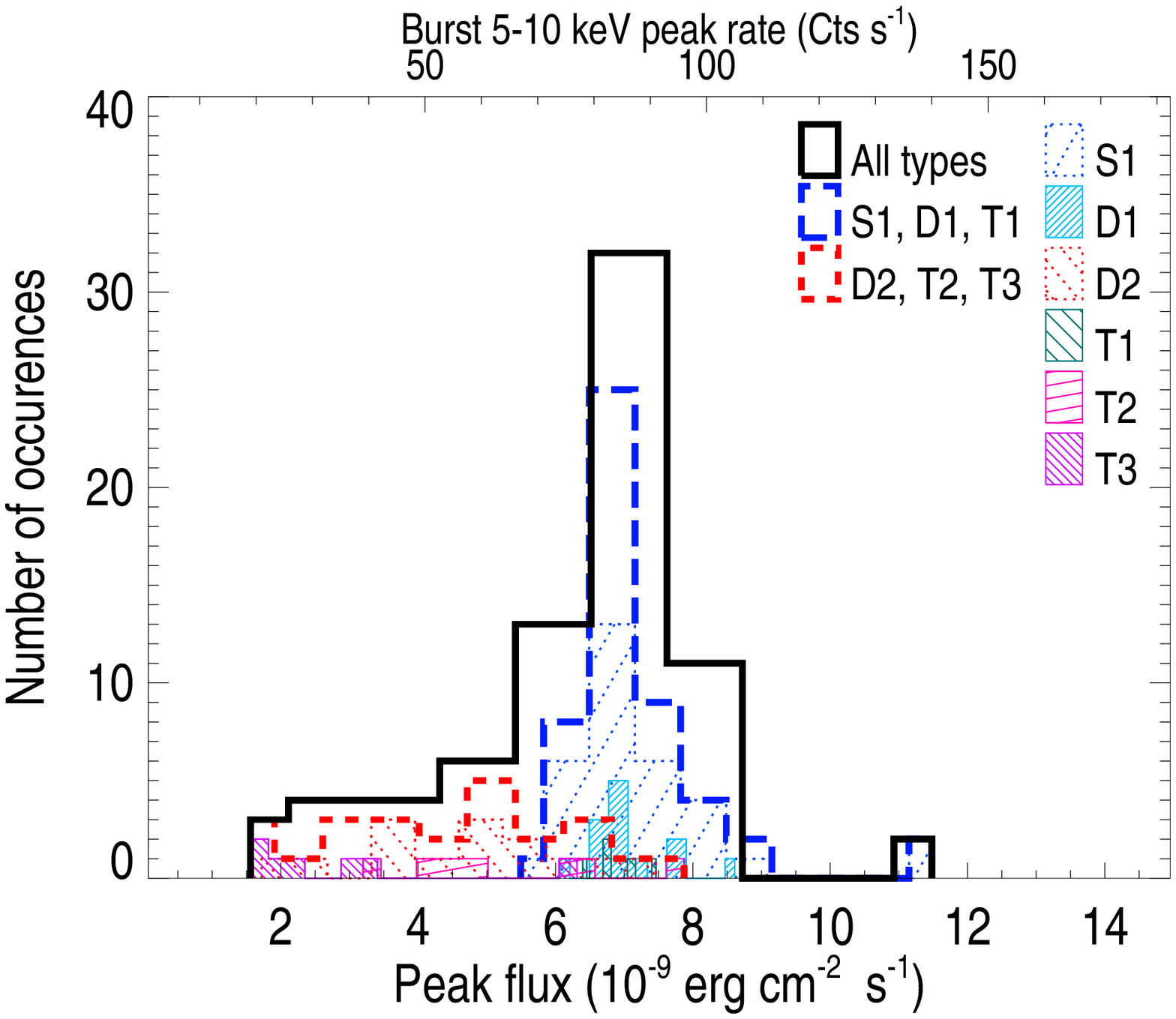}}
%\centerline{\includegraphics[angle=0,width=0.45\textwidth]{fig_histo_fluence.ps}}
\centerline{\includegraphics[angle=0,width=0.45\textwidth]{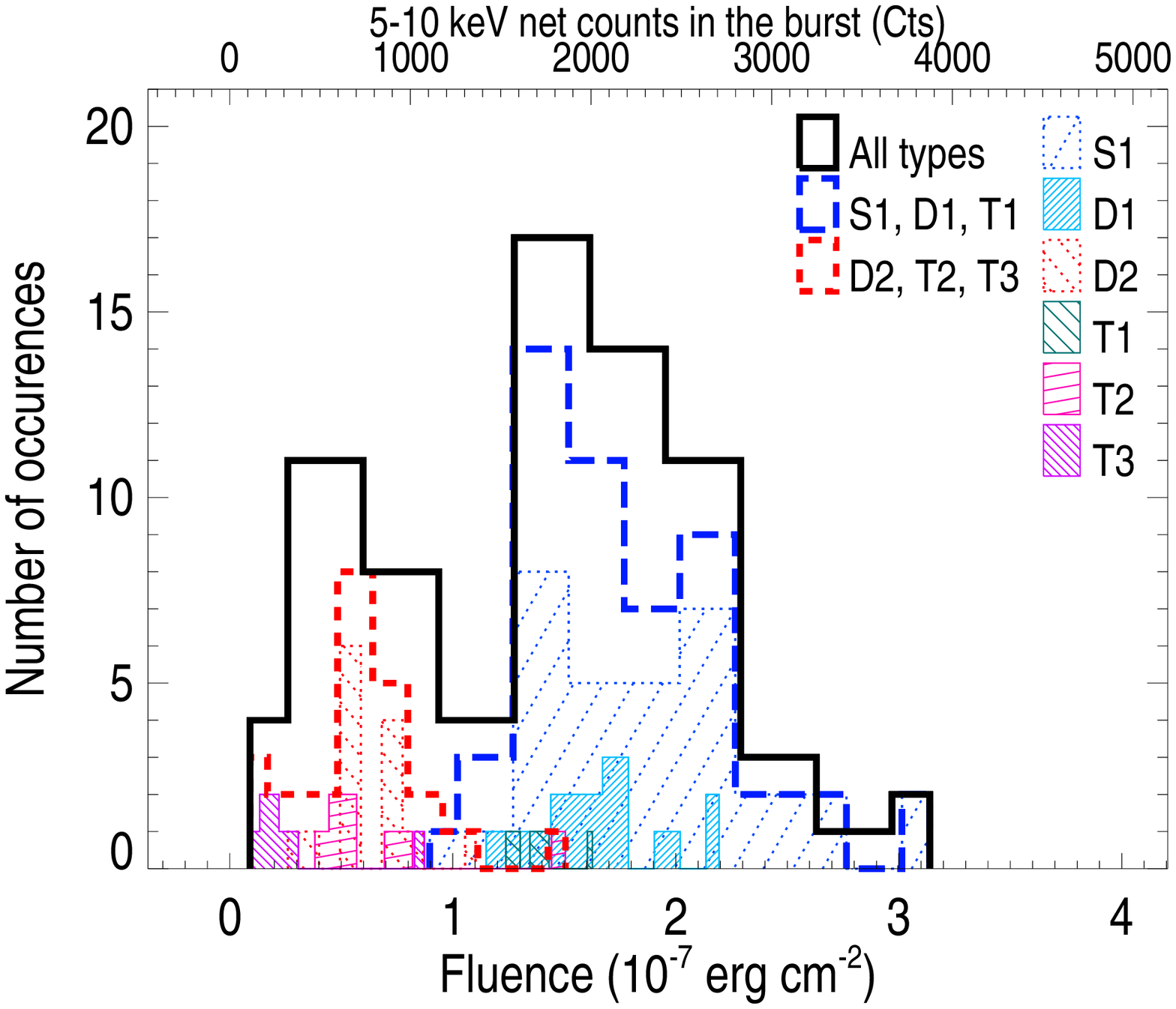}}
\vspace{-0.5cm}
%\centerline{\includegraphics[width=0.45\textwidth]{fig_histo_alpha.ps}}
\centerline{\includegraphics[width=0.45\textwidth]{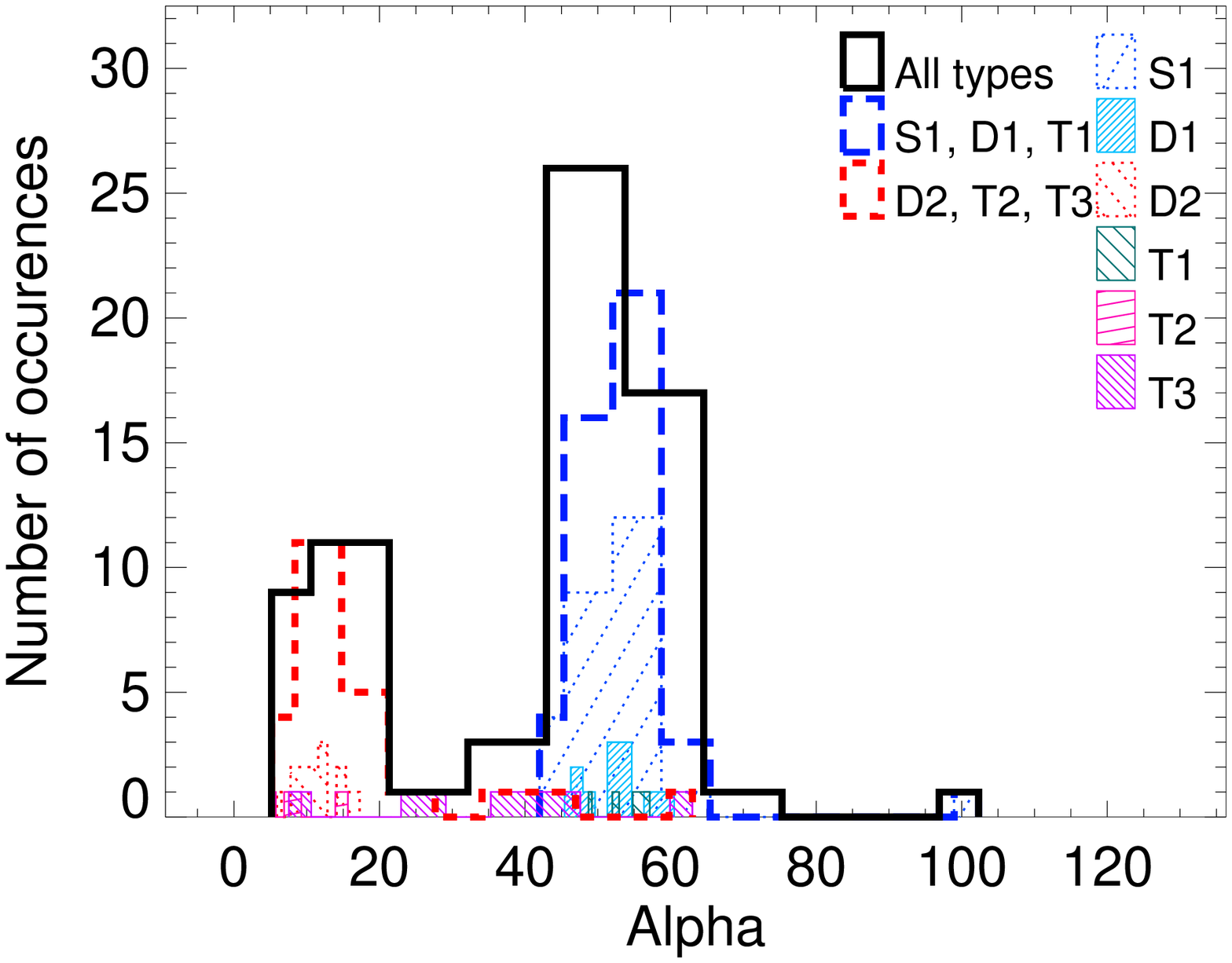}}
\caption{Histograms of the burst peak flux (top), fluence (middle) and $\alpha$ values (bottom) built as  in Fig.~\ref{fig:wait}. }
\label{fig:histo}
\end{figure}

Fig.~\ref{fig:waitnextevent} shows the histograms of the event wait
time to the next event. While the doublets and the triplets seem to
follow the same distribution (\psame{doublet-triplet}~=~58\%), the
singlets follow a significantly different distribution than that of
the doublets and triplets (\psame{singlet-doublet}~=~0.0015\% and
\psame{singlet-triplet}~=~0.2\%). For a new event to occur, one has to
wait on average 4.2~h after a triplet, 3.7~h after a doublet, but only
2.7~h after a singlet.

\subsubsection{Peak flux}

The histograms of the burst peak flux are shown in
Fig.~\ref{fig:histo} (top; see also Table~\ref{tab:mean}).  Despite
their overlap, the LWT and SWT peak flux histograms are inconsistent
with coming from the same distribution (\psame{long-short}~=~$1.5
\times 10^{-8}$). The mean peak flux of the LWT bursts is a factor 1.6
higher than that of the SWT bursts.  The mean peak flux of the LWT
bursts, $7.2 \times 10^{-9}$~\ergcms, corresponds to a luminosity of
$2.2 \times 10^{37}$~\ergs\ at 5~kpc or $8.6 \times 10^{37}$~\ergs\ at
10~kpc. The mean peak flux of the SWT bursts, $4.5 \times
10^{-9}$~\ergcms, corresponds to a luminosity of $1.3 \times
10^{37}$~\ergs\ at 5~kpc or $5.4 \times 10^{37}$~\ergs\ at 10~kpc.

 The LWT peak flux histogram is very peaked (kurtosis of 5.68) while
that of the SWT group is flat (kurtosis of -1.06).  The S1, D1 and T1
bursts have similar peak flux distributions (\psame{S1-D1}~=~84\%,
\psame{S1-T1}~=~96\%, \psame{D1-T1}~=~99\%). In the SWT group, the D2
and T2 bursts have similar peak flux distributions
(\psame{D2-T2}~=~99\%), while the T3 bursts seem to follow a
marginally different one (\psame{D2-T3}~=~4.5\%,
\psame{T2-T3}~=~3.6\%), with lower values.

In a given doublet (triplet), the peak flux of the D2 (T2) is
significantly lower than that of the D1 (T1) in 10/13 (3/5) cases, and
similar to it in the other cases (see Fig.~\ref{fig:burst_profiles}).
The peak flux of the T3 burst is always significantly less than that
of the T2 burst, except for the strong T3 burst number 24.

\subsubsection{Fluence and duration}

The histograms of the burst fluence are shown in Fig.~\ref{fig:histo}
(middle; see also Table~\ref{tab:mean}). The fluence sample is clearly
bimodal (\psame{long-short}~=~7.6~$\times 10^{-14}$), the SWT bursts
being less fluent by a factor $\sim$3 than the LWT bursts on average.
The mean fluence of the LWT bursts is $1.8 \times 10^{-7}$~\ergcm\ and
corresponds to $5.3 \times 10^{38}$~ergs at 5~kpc or $2.1 \times
10^{39}$~ergs at 10~kpc.  The mean fluence of the SWT bursts is $5.7
\times 10^{-8}$~\ergcm\ and corresponds to $1.7 \times 10^{38}$~ergs
at 5~kpc or $6.8 \times 10^{38}$~ergs at 10~kpc.

\begin{figure*}[!t] 
%\centerline{\hspace{0.5cm}\includegraphics[angle=0,height=0.235\textheight]{fig_flu_accu_comb.eps}\hspace{-1cm}\includegraphics[angle=0,height=0.235\textheight]{fig_fluence_accu_group_color.ps}}
\centerline{\hspace{0.5cm}\includegraphics[angle=0,height=0.235\textheight]{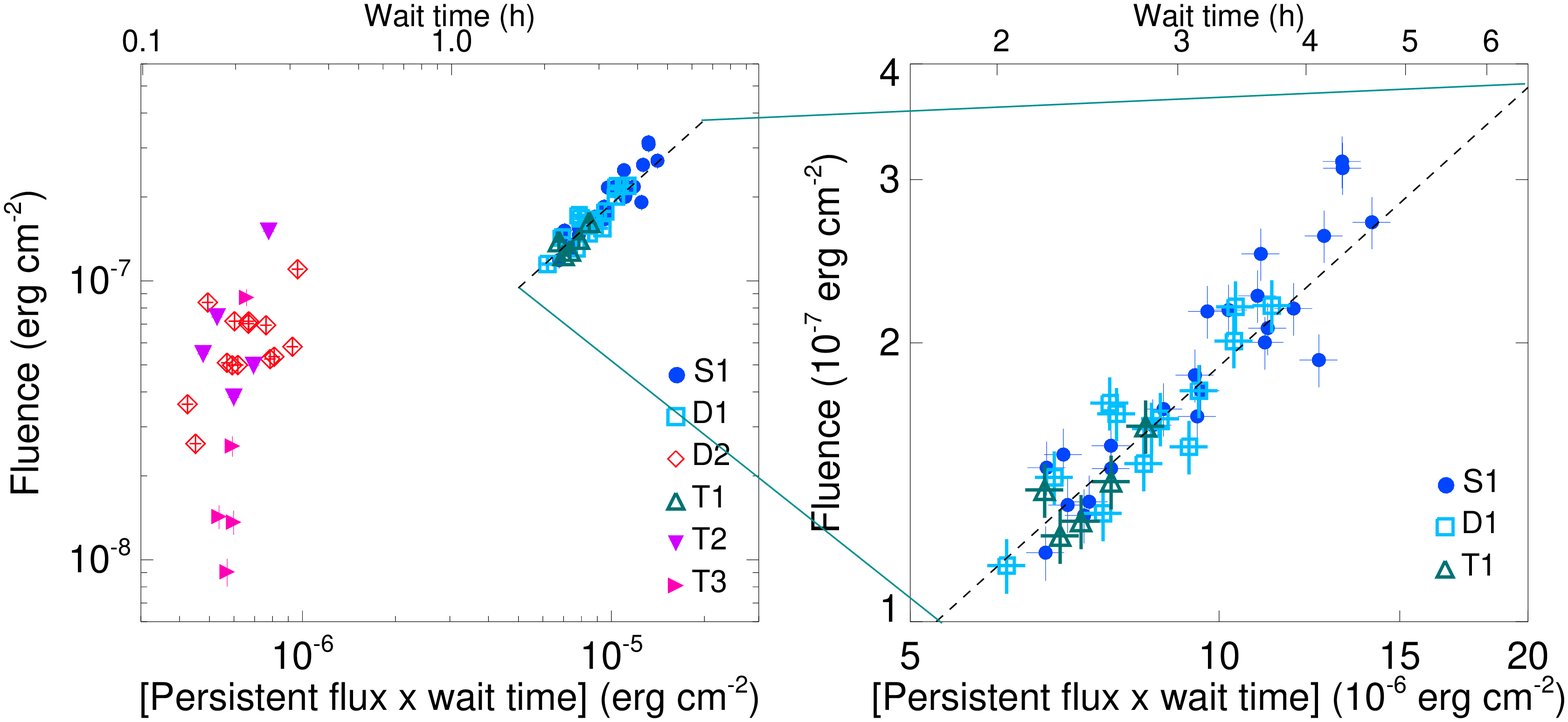}\hspace{-1cm}\includegraphics[angle=0,height=0.235\textheight]{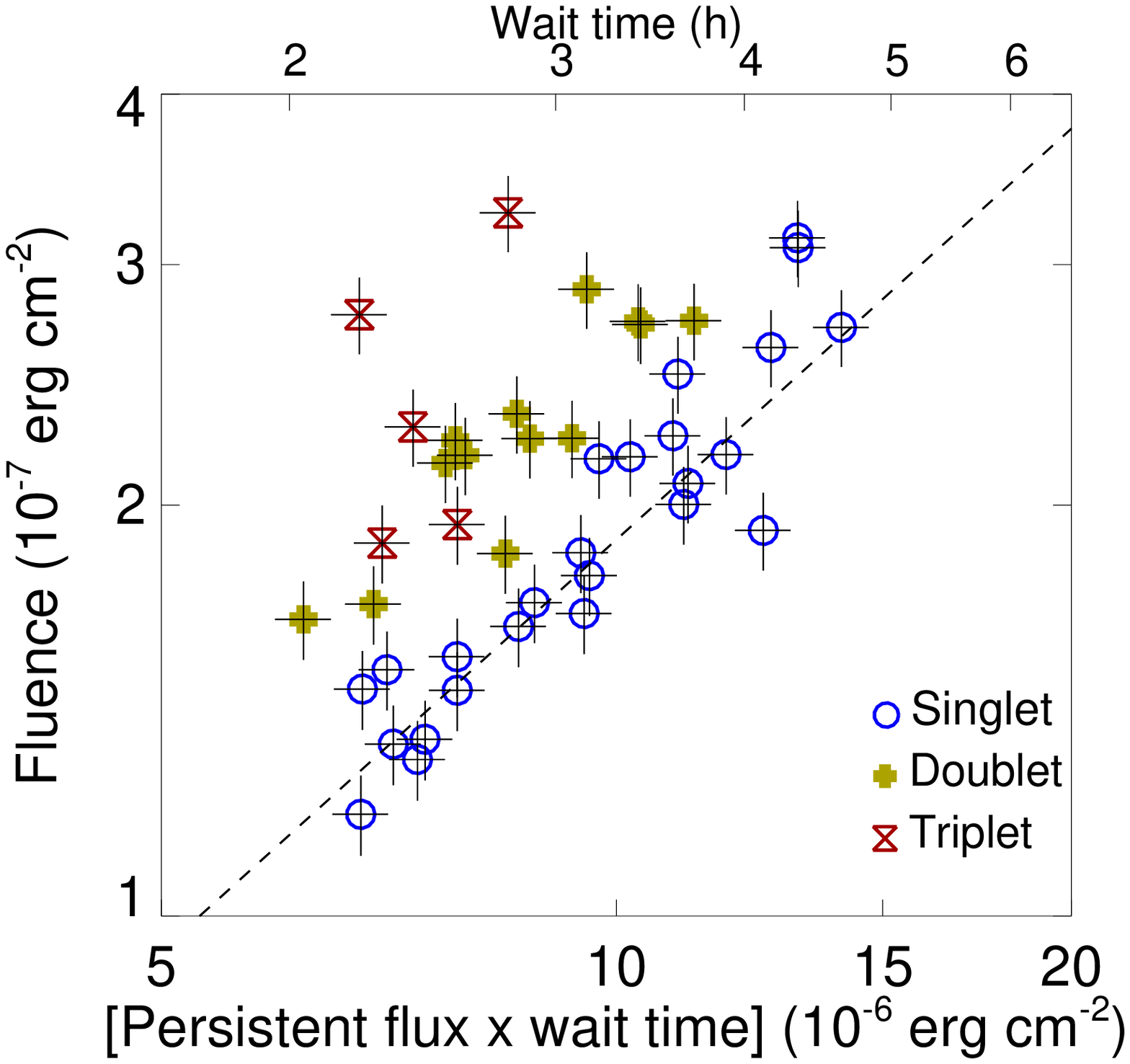}}
\caption{{\bfseries Left:} Burst fluence as a function of persistent
flux times wait time (accumulated mass tracer).  The line indicates
the mean $\alpha$ value for the LWT bursts ($\alpha$~=~53).
{\bfseries Middle:} zoom in on the left panel. {\bfseries Right:}
Event fluence as a function of the accumulated mass tracer.  The line
is the same as in the left panel.}
\label{fig:fluencewait}
\end{figure*}

Inside the LWT group, the fluences from the S1 and D1 bursts seem to
be drawn from the same distribution (\psame{S1-D1}~=~26\%), while the
T1 bursts would follow a marginally different one
(\psame{S1-T1}~=~4.1\%, \psame{D1-T1}~=~5.4\%), the T1 being on
average less fluent than the D1 or the S1 bursts.  Inside the SWT
group, the D2 and T2 have similar fluence distributions
(\psame{D2-T2}~=~93\%), while the T3 have marginally different (lower)
ones (\psame{D2-T3}~=~0.7\%, \psame{T2-T3}~=~3.6\%).  The histograms
of the burst duration (not shown here) are very similar to the fluence
ones.

In a given doublet, the D2 burst is always shorter and less fluent
than the D1. In a given triplet, the duration and fluence of the T2 is
always less or equal to that of the T1. The T3 burst, except for the
strong T3 burst number 24, has a duration and fluence less than that
of the T2 burst.

\subsubsection{Alpha}

The histograms of the $\alpha$ values are shown in
Fig.~\ref{fig:histo} (see also Table~\ref{tab:mean}). The sample is
clearly bimodal (\psame{long-short}~=~1.0~$\times 10^{-12}$), the mean
$\alpha$ value being smaller by a factor $\sim$3.3 for the SWT bursts
($\alpha = 16$) than for the LWT ones ($\alpha = 53$).  The $\alpha$
histogram for the S1 bursts is very peaked (kurtosis of 11.4) compared
to all the other burst types (kurtosis between -0.4 and -2).  Inside
the LWT group, the S1, D1 and T1 bursts seem to follow the same
$\alpha$ distribution (\psame{S1-D1}~=~90\%, \psame{S1-T1}~=~48\%,
\psame{D1-T1}~=~60\%).  Inside the SWT group, the T3 bursts seem to
follow a marginally different $\alpha$ distribution from the D2 and T2
bursts (\psame{D2-T3}~=~1.9\%, \psame{T2-T3}~=~3.6\% while
\psame{D2-T2}~=~32\%).

\subsection{Relation between burst fluence and  accumulated mass}
\label{sec:massaccu}

Fig.~\ref{fig:fluencewait} (left and middle panels) shows the burst
fluence as a function of the product of the persistent flux times the
wait time, which is a tracer of the mass accumulated by accretion
before the burst.  Here, the accumulated mass tracer is simply
proportional to the wait time because we assume the underlying
persistent flux is constant (Sect.~\ref{sec:nonbursting}).

Among the LWT types, the fluence is strongly correlated with the
accumulated mass tracer, with correlation coefficients of 0.91 for the
S1, 0.89 for the D1 and 0.81 for the T1 bursts.  For each LWT type,
the fluence is well described by a linear function of the accumulated
mass tracer ($E_{\rm b} = \alpha^{-1} \times [ f_{\rm p} \times t_{\rm
wait}]+ E_0$). $E_0$ is the offset energy at zero burst interval.  The
best-fit parameter values $\alpha$ and $E_0$ are listed in
Table~\ref{tab:fitfluenceaccu}. Within the 90\% confidence range, both
$\alpha$ and $E_0$ are consistent with being the same for the S1, D1
and T1 bursts.  For all the LWT bursts together, the $\alpha$ value
obtained from the fit is $46\pm4$ and $E_0$ is $ (-2.3 \pm 1.7) \times
10^{-8}$~\ergcm. $E_0$ is consistent with 0 within 3$\sigma$, and
consistent with being negative within the 90\% confidence range
(1.65$\sigma$).  A negative value of $E_0$ would indicate that a part
$\left| E_0 \right| = 2.3 \times 10^{-8}$~\ergcm\ of the total
available energy is not burned in the burst. This corresponds to an
energy of $6.9 \times 10^{37}$~ergs at 5~kpc or $ 2.7 \times
10^{38}$~ergs at 10~kpc, equivalent to 13\% of the mean energy
released by the LWT bursts.

\begin{table}[!b]
\caption{Best-fit parameters $\alpha$ and $E_0$, and their 90\% confidence
 errors, obtained from fitting $E_{\rm b}$ as $\alpha^{-1} \times [
 f_{\rm p} \times t_{\rm wait}] + E_0$ (see text).}
\begin{center}
\begin{tabular}{l c c }
\hline 
\hline 
\noalign {\smallskip}
 Type & $\alpha$ & $E_0$ ($10^{-8}$ \ergcm) \\
\hline 
\noalign {\smallskip}
S1 & $  45 \pm 5 $ & $  -2.5 \pm 2.3 $ \\
D1 & $   49 \pm 11$ & $  -1.1 \pm 3.7 $\\
T1 &  $   50 \pm 40 $ & $  1 \pm 13 $ \\
S1, D1, T1 &  $   46 \pm 4 $ &$ -2.3 \pm 1.7$ \\
\hline 
\noalign {\smallskip}
\end{tabular}
\end{center}
\label{tab:fitfluenceaccu}
\end{table}

Among the SWT types, only the T3 bursts show a strong correlation
(coefficient of 0.86) between the fluence and the accumulated mass,
which can be described by a line with best-fit parameters $\alpha =
1.2 \pm 0.4 $ and $E_0 = (-4.5 \pm 4.6) \times 10^{-7}$~\ergcm\ (90\%
confidence errors). The fluence of the D2 or T2 bursts is much less
strongly correlated with the accumulated mass, the coefficients being
0.53 for the D2 and 0.62 for the T2, and no acceptable linear fit
could be obtained. So, apart maybe for the T3 bursts, the fluence of a
SWT burst does not strongly depend on the amount of material freshly
accreted just before the SWT burst.

Combining SWT and LWT types (e.g. the whole sample of bursts, or the
D1 and D2 combined), we could not obtain acceptable (with reduced
\chisq\ $<$2) linear fits between the fluence and the accumulated mass
tracer.  Thus, the SWT bursts are inconsistent with following the same
fluence- accumulated mass relation as the LWT bursts.

Fig.~\ref{fig:fluencewait} (right) shows the {\em event} fluence as a
function of the accumulated mass tracer.  At a given accumulated mass,
a triplet is generally more fluent than a doublet which is more fluent
than a singlet.  When all the events are considered together, the
event fluence is only weakly correlated to the accumulated mass tracer
(correlation coefficient of 0.38).  The fluence of the doublets
correlates with the accumulated mass.  However, this correlation is
not better than the one observed for the D1 bursts only.  Similarly
the correlation is not improved by considering the triplets
(coefficient of 0.44) rather than the T1 bursts (coefficient of 0.81).
This effect is visible when the left and right panels of
Fig.~\ref{fig:fluencewait} are compared: the dispersion is higher for
the doublets than for the D1 and higher for the triplets than for the
T1.  This suggests that the fluence of a D2 burst (and analogously of
a T2-T3 pair) does not strongly depend on the amount of material
accreted before the D1 (or T1) burst of the considered event.

\subsection{Relation between fluence and flux}

\begin{figure}[!t] 
%  \centerline{\includegraphics[angle=0,width=0.45\textwidth]{fig_fluence_flux.ps}}
  \centerline{\includegraphics[angle=0,width=0.45\textwidth]{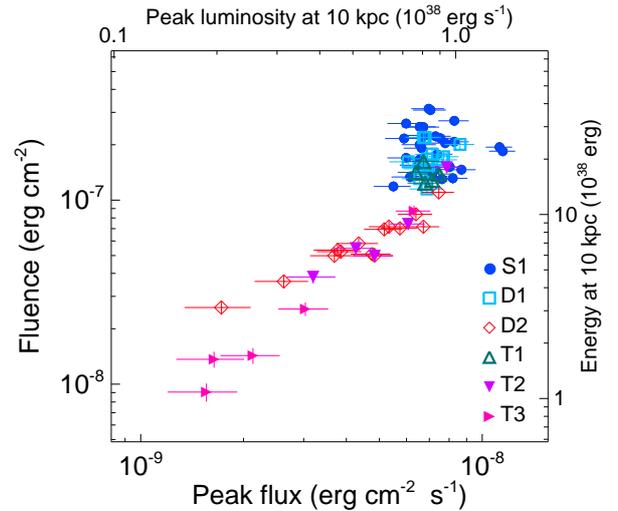}}
  \caption{Bolometric burst fluence as a function of the bolometric
  peak flux. The right and top axis indicate the corresponding energy
  and peak luminosity assuming an isotropic emission at a
  distance of 10~kpc.}
\label{fig:fluenceflux}
\end{figure}

Fig.~\ref{fig:fluenceflux} shows the burst fluence as a function of
the burst peak flux. While the two quantities are not strongly
correlated for the S1, D1 and T1 bursts (with linear correlation
coefficients of 0.04, 0.29, and -0.27, respectively), the fluence and
flux are strongly correlated in the case of the D2, T2 and T3 bursts,
with correlation coefficients of 0.93, 0.94 and 0.99, respectively.

\subsection{Profiles} 
\label{sec:profiles}

 \begin{figure}[!t]
% \centerline{\includegraphics[angle=90,width=0.45\textwidth]{fig_averagedprofiles_all_bin.ps}}
 \centerline{\includegraphics[angle=90,width=0.45\textwidth]{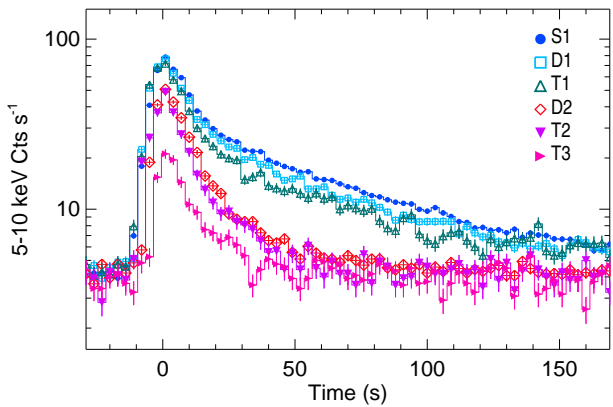}}
\caption{Average profiles for each burst type, obtained respectively
from 32 S1, 13 D1, 14 D2, 5 T1, 4 T2 and 4 T3 bursts. The strong T2
and T3 bursts number 35 and 24 are excluded. The binning is 3~s. Note
that the logarithmic scale of the intensity axis makes an exponential
decay appear as a straight line.}
\label{fig:aveprofiles}
\end{figure}

Fig.~\ref{fig:aveprofiles} shows the profiles obtained by averaging
the 5--10~keV light curves of all the bursts of a given type, except
the exceptionally strong T2 and T3 bursts number 35 and 24.
After subtracting the persistent count rate level before
the bursts, the decay of all these profiles is well described by a
model consisting of two exponential parts. The best-fit values of the
characteristic decay times, $\tau_1$ and $\tau_2$, and of the
transition time $t_1$ are given in Table~\ref{tab:fitprofiles}.  The
first part of the decay is always faster than the second
(\taua$<$\taub).  The decays of the average S1, D1 and T1 profiles are
similar to each other, with \taua~$\sim$~15~s and
\taub~$\sim$~52~s. The D2 and T2 average profiles decay more rapidly,
with both \taua\ and \taub\ being shorter.  For the T3 average
profile, only one exponential component is detected (including a
second one does not improve the fit quality). Its decay time is
similar to that of the first component of the D2 and T2 average
profile.

\begin{figure}[!t]  
%\centerline{\includegraphics[angle=90,width=0.45\textwidth]{fig_averagedprofiles_s1d1t1d2_samepeakwait.ps}}
\centerline{\includegraphics[angle=90,width=0.45\textwidth]{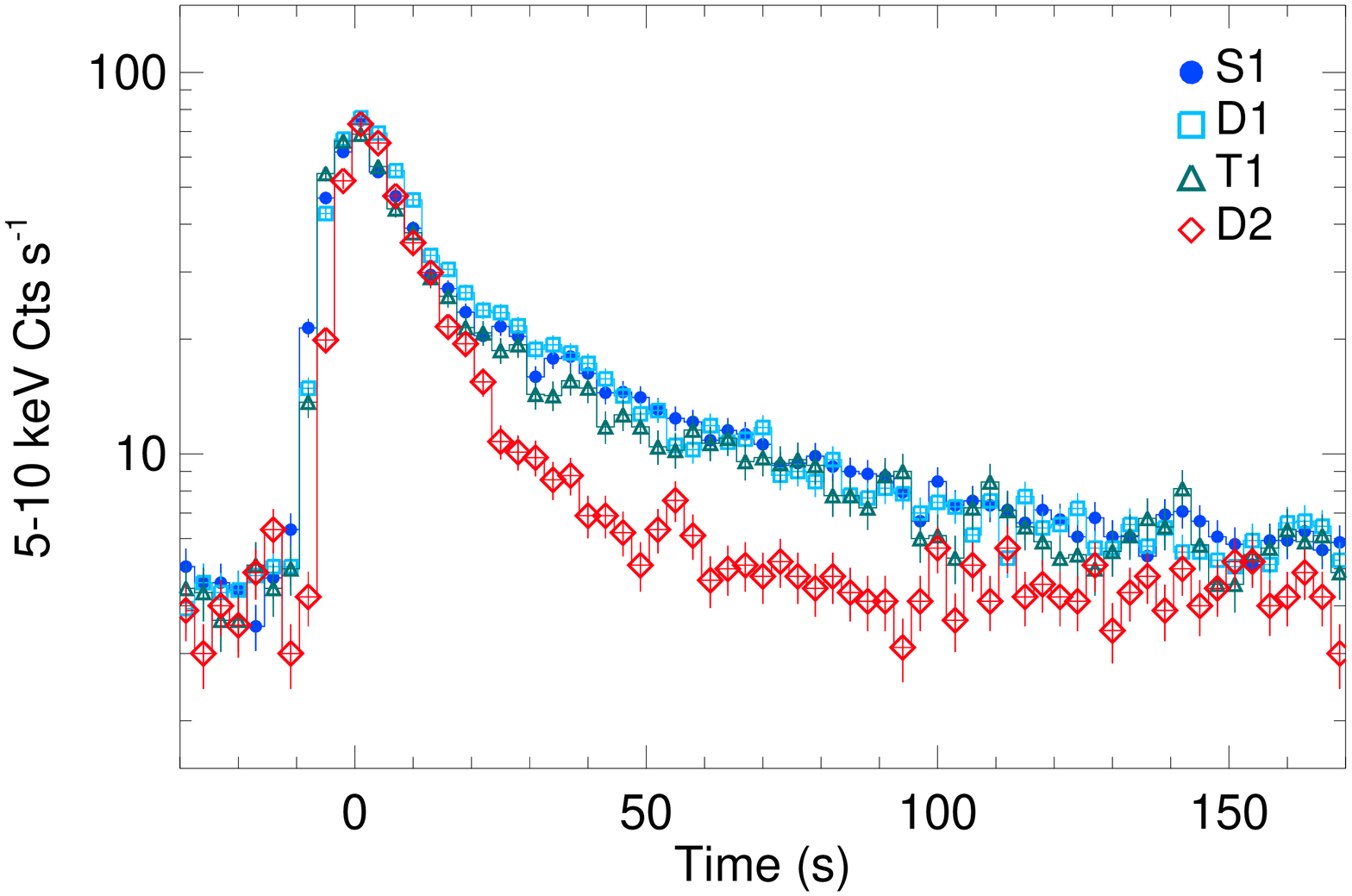}}
%\centerline{\includegraphics[angle=90,width=0.45\textwidth]{fig_averagedprofiles_samepeak_s1d2strongt3.ps}}
\centerline{\includegraphics[angle=90,width=0.45\textwidth]{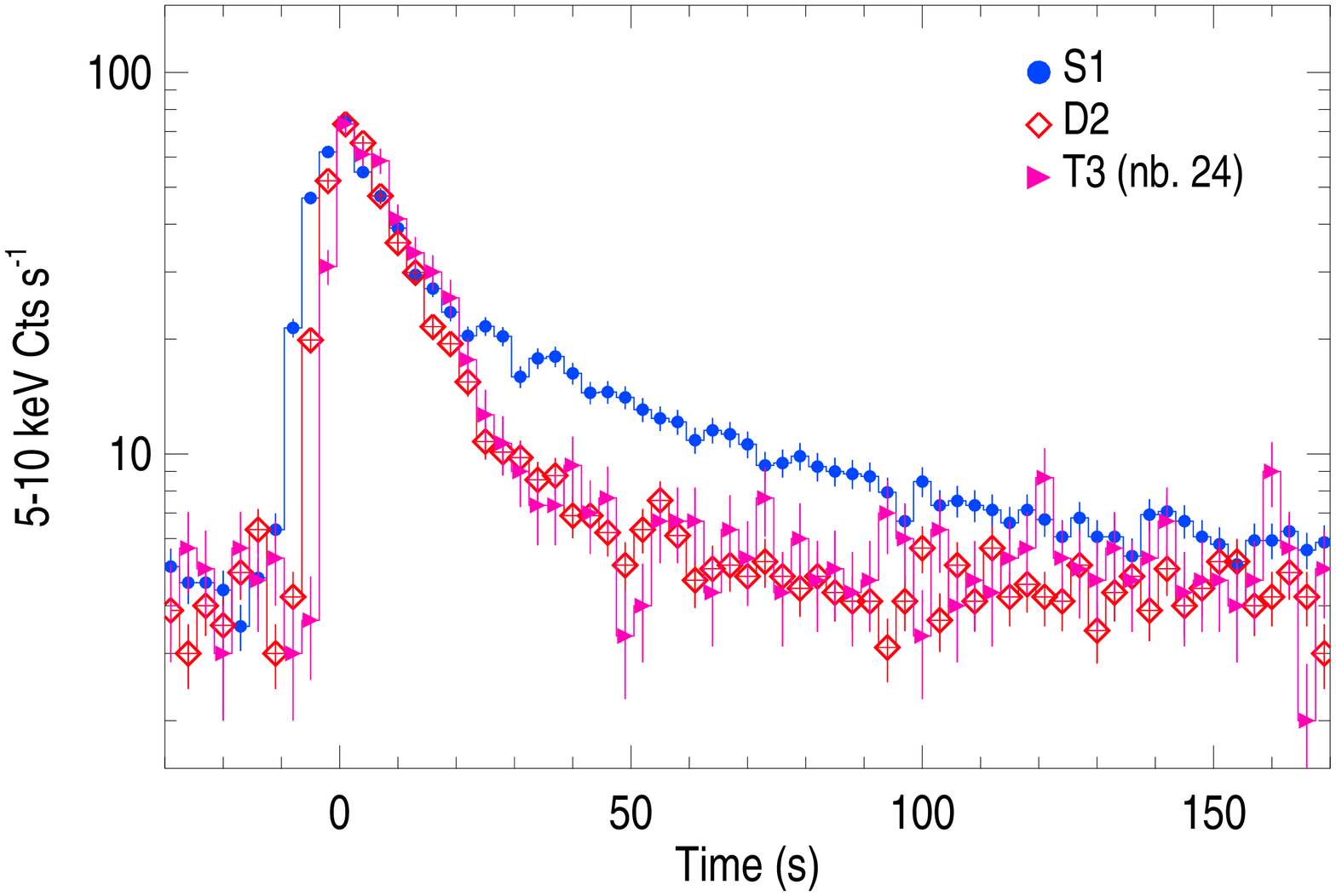}}
\caption{{\bfseries Top:} Average S1, D1, T1 and D2 profiles, obtained
respectively from 5~S1, 5~D1, 3~T1 and 3~D2 bursts having a similar
peak intensity (77--96~\countsec), and a similar wait time for the S1,
D1 and T1 (2--2.8~h). Time is given from the burst peak time and the
binning is 3~s. {\bfseries Bottom:} The strong T3 number 24 compared
with the S1 and D2 average profiles from the top panel.}
\label{fig:profilessamepeak}
\end{figure}

The top panel of Fig.~\ref{fig:profilessamepeak} shows the average S1,
D1, T1 and D2 profiles obtained from bursts having a similar peak
intensity in the range 77--96~\countsec, and a similar wait time in
the range 2--2.8~h for the S1, D1 and T1. The best-fit model of their
decay with two exponential parts is given in
Table~\ref{tab:fitprofilessamepeak}.  While the S1, D1 and T1 profiles
are similar, the D2 profile, {\em at the same peak intensity level},
shows a different and much faster decay which can be modeled by only
one exponential decay.

The only T2 and T3 bursts that have a peak intensity in the range
77--96~\countsec\ are the exceptionally strong ones number 35 and 24,
respectively.  The best-fit model of their decay is given in
Table~\ref{tab:fitprofilessamepeak}. The decay of the strong T2 burst
number 35 shows a fast part followed by a slow flat component.
Interestingly, the decay of the strong T3 burst (bottom panel of
Fig.~\ref{fig:profilessamepeak}) is very similar to that of the
average D2 profile {\em at the same peak intensity level}.

\begin{table*}[!t]
\caption{Best-fit parameter values and 90\% confidence errors for the
decay of the average profiles by burst type
(Fig.~\ref{fig:aveprofiles}).  We subtract the persistent
count rate level before the burst and fit each persistent-subtracted
burst decay using a continuous function of two exponential parts: $B_0
e^{-t/ \tau_1}$ between $t_0$ and $t_1$, and $B_1 e^{-t/\tau _2}$
between $t_1$ and $t_2$. We set $t_0$ to the burst peak time (0~s),
$t_2$ to the first time where the persistent-subtracted intensity,
$B$, reaches a level of 0, and $B_1$ to $B_0 e^{(-t_1/ \tau_1)}$.  The
free parameters are the persistent-subtracted intensity, $B_0$, at the
burst peak time, the characteristic decay times, $\tau_1$ and
$\tau_2$, and the transition time $t_1$.  We indicate the time,
$t_{\rm stop}$, where the intensity, $I$, reaches back a level of 1.1
times the level before the burst. For the profiles whose fit was not
improved by including a second exponential component, we give only the
decay time of the first and only exponential component.}
\begin{center}
\begin{tabular}{l c c c  c c}
\hline 
\hline 
 Type & $B_0$ & \taua & \taub & \ta & $t_{\rm stop}$\\
  & (\countsec) & (s) & (s)  & (s) & (s)\\
\hline 
\noalign {\smallskip}
S1 & $79.7^{+1.7}_{-0.5}$ & $16.2^{+0.4}_{-0.1}$ & $56.9^{+2.1}_{-0.6}$ & $20.7^{+0.2}_{-0.7}$ & $180$ \\
D1 & $74.1^{+3.0}_{-0.8}$ & $15.9^{+0.6}_{-0.2}$ & $51^{+4}_{-1}$ & $21.6^{+0.4}_{-1.5}$ & $126$ \\
T1 & $71.1^{+4.8}_{-1.5}$ & $13.8^{+1.0}_{-0.3}$ & $48^{+12}_{-2}$ & $20.6^{+0.6}_{-2.3}$ & $103$ \\
D2 & $53.9^{+2.2}_{-1.5}$ & $11.0^{+0.6}_{-0.2}$ & $19^{+13}_{-2}$ & $25^{+2}_{-5}$ & $52$ \\
T2 & $50.5^{+4.6}_{-2.7}$ & $9.0^{+1.1}_{-0.4}$ & $14^{+10}_{-1}$ & $15^{+3}_{-5}$ & $45$ \\
T3 & $17.3^{+8.2}_{-0.4}$ & $10.8^{+8.7}_{-0.2}$ & --- & --- & $33$ \\
\hline 
\noalign {\smallskip}
\end{tabular}
\end{center}
\label{tab:fitprofiles}
\end{table*}

\subsection{Spectral properties}

\subsubsection{Color intensity diagram}

\begin{table*}[!t]
\caption{Same as Table~\ref{tab:fitprofiles} for the average profiles
by type from bursts having a peak intensity in the range
77--96~\countsec\ and a wait time in the range 2--2.8~h for the S1, D1
and T1 (Fig.~\ref{fig:profilessamepeak}).}
\begin{center}
\begin{tabular}{l c c c  c c}
\hline 
\hline 
 Type & $B_0$ & \taua & \taub & \ta & $t_{\rm stop}$\\
  & (\countsec) & (s) & (s)  & (s) & (s)\\
\hline 
\noalign {\smallskip}
S1 & $70^{+5}_{-1}$ & $14.0^{+1.1}_{-0.2}$ & $49^{+10}_{-2}$ & $19.9^{+0.5}_{-2.6}$ & $116$ \\
D1 & $81^{+6}_{-1}$ & $14.2^{+1.2}_{-0.2}$ & $40^{+8}_{-1}$ & $20.3^{+0.5}_{-3.0}$ & $105$ \\
T1 & $68^{+8}_{-2}$ & $13.7^{+1.9}_{-0.2}$ & $45^{+24}_{-2}$ & $21.0^{+0.6}_{-4.6}$ & $101$ \\
D2 & $75^{+10}_{-1}$ & $11.6^{+1.5}_{-0.2}$ & --- & --- & $50$ \\
T2 & $78^{+22}_{-2}$ & $13.0^{+3.5}_{-0.2}$ & $149^{<-277~a}_{-18}$ & $30.7^{+0.6}_{-9.0}$ & $108$ \\
T3 & $92^{+8}_{-6}$ & $10.8^{+2.3}_{-0.3}$ & --- & --- & $37$ \\
\hline 
\noalign {\smallskip}
\multicolumn{6}{l}{$^{a}$ The second component of this decay is effectively flat,}\\
\multicolumn{6}{l}{which shows in our fits as a long positive or negative exponential decay time.}
\end{tabular}
\end{center}
\label{tab:fitprofilessamepeak}
\end{table*}

\begin{figure}[!t] 
%  \centerline{\includegraphics[width=0.5\textwidth]{fig_decay_color_intensity.ps}}
  \centerline{\includegraphics[width=0.5\textwidth]{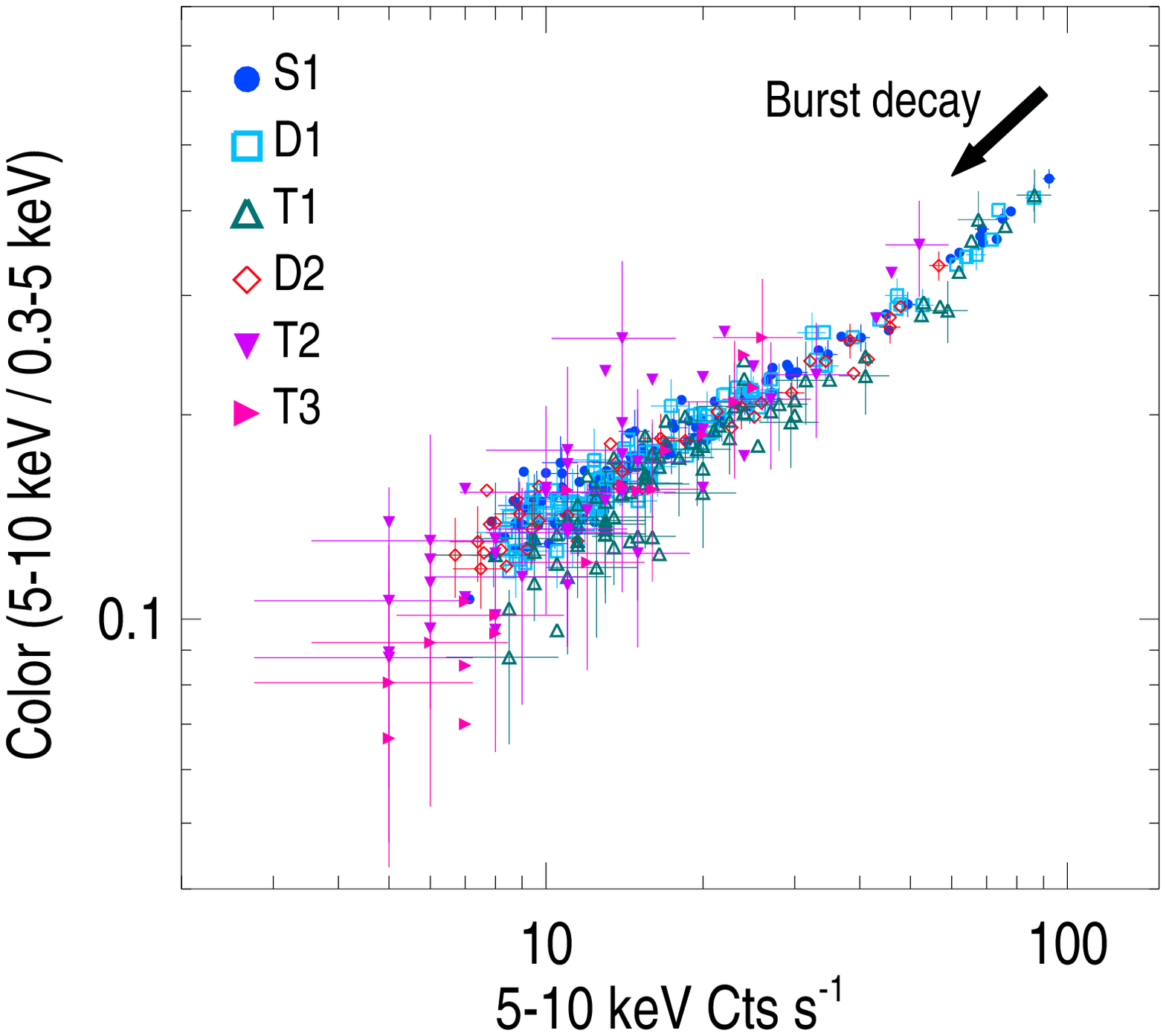}}
\caption{Color as a function of hard intensity during the decay of the
averaged profiles by type obtained, respectively, from 15~S1, 8~D1,
10~D2, 2~T1, 1~T2 and 1~T3 bursts which are the least affected by the
dipping activity. Error bars are shown on one third of the points
which represent each 1~s.}
\label{fig:avehrdecay}
\end{figure}

Fig.~\ref{fig:avehrdecay} shows the color, $C$, as a function of the
hard intensity, $I$, during the decay of the averaged profiles
obtained, respectively, from 15~S1, 8~D1, 10~D2, 2~T1, 1~T2 and 1~T3
bursts chosen for being the least contaminated by the dipping
activity.  We find that, for each burst type, $\log(C)$ can be
described by a linear function of $\log(I)$ ($\log(C) = a \times
\log(I) + b$).  For a burst emitting like a blackbody, the bolometric
flux $F$ is expected to depend only on the area, $A$, of the emitting
region and on the blackbody temperature, $T$, as $F \propto A T^4$.
Since $C$ is a tracer of $T$ and $I$ is a tracer of $F$, any
difference in the slope $a$ between burst types would indicate a
difference in the flux to temperature dependence ($F \propto T^4$),
while any difference in the intercept $b$ would indicate a difference
in the area $A$ of the emitting region.  We find that, within
$3\sigma$, the slope $a$ is consistent with being the same for all
burst types. So the comparison of the averaged color-intensity
diagrams by burst types does not indicate differences by type in the
burst temperature to flux dependence.  
The best-fit values of the intercept $b$ are consistent with being the
same within $3\sigma$ for the S1, D1, T1, T2 and T3 types on one hand,
and for S1, D1, D2, T2 and T3 types on the other hand, indicating a
difference only between the T1 and D2 types. This difference suggests
that the apparent emitting area (or blackbody radius) would be on
average larger for the T1 than for the D2 bursts.

\begin{figure}[!t] 
%  \centerline{\includegraphics[width=0.5\textwidth]{fig_peak_color_intensity.ps}}
  \centerline{\includegraphics[width=0.5\textwidth]{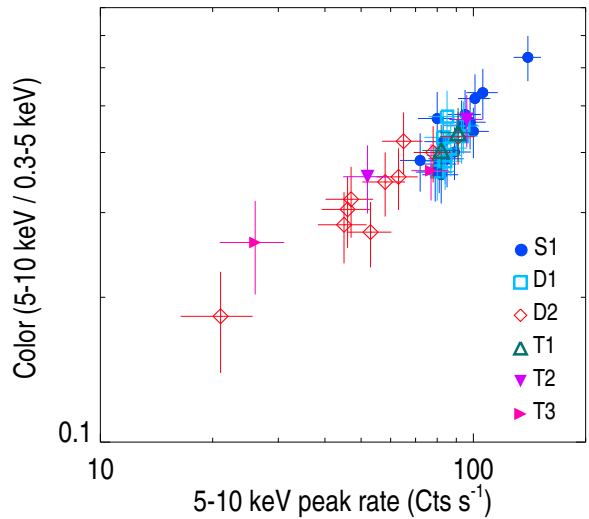}}
\caption{Color as a function of the hard intensity at the burst peak
for the individual bursts the least affected by dipping activity.}
\label{fig:avehrpeak}
\end{figure}

Fig.~\ref{fig:avehrpeak} shows the color at the burst peak as a
function of the intensity at the peak, for the individual bursts the
least affected by dipping activity.  The two quantities are strongly
correlated.  This implies that the peak temperature distributions have
similar properties as the peak fluxes distributions shown in
Fig.~\ref{fig:histo}, with a clear distinction between the SWT and the
LWT bursts, the latter having higher and less variable peak
temperatures.  In addition, the correlation in
Fig.~\ref{fig:avehrpeak} is consistent with that obtained from the
averaged decays (Fig.~\ref{fig:avehrdecay}). So the relation between
the temperature and the flux is consistent with being the same for all
burst types and for any time of the burst decay including at the peak.

This analysis shows that there is no indication for strong spectral
differences between the various burst types, the flux or apparent
emitting area being approximately the same at a given temperature
during the decay. There is however marginal evidence that the apparent
emitting area is on average larger for the first bursts in triplets
than for the second bursts in doublets. This could point to
differences in the structure or composition of the neutron star
atmosphere/outer layers (see Sect.~\ref{sec:exosat} and the discussion
in \citealt{0748:gottwald87apj}).

\subsubsection{Time-resolved spectral modeling}
\label{sec:spectral}

We select five singlets (bursts 1, 12, 16, 33 and 37, see Appendix),
one doublet (bursts 4 and 5) and two triplets (bursts 22, 23, 24 and
34, 35, 36) that are the least contaminated by dipping activity and
perform their time-resolved spectral analysis using version 11.3.1 of
the XSPEC package.  We divide each burst into 3 to 6 intervals, with
the interval around the burst peak chosen shorter than intervals in
the tail. We extract a spectrum in each burst interval from which we
subtract a reference persistent spectrum obtained from segments just
before and after the burst.  We add a 2\% systematic error to account
for calibration uncertainties.

Since the net emission of Type-I X-ray bursts is generally well
described by a blackbody \citep[see e.g. ][]{swank77apjl,lewin93ssr},
we model the net 0.1--10~keV spectrum with an absorbed blackbody.  We
fix the hydrogen column density, $N_{\mathrm{H}}$, of the neutral
absorber to $0.1\times 10^{22}\,\mathrm{cm^{-2}}$
\citep[][]{0748:sidoli05aa}.  We also include a Gaussian with a
centroid energy set to 0 to describe emission below $\sim$2~keV which
is in excess of the blackbody model.  This soft excess may be related
to a temporary increase, due to the burst irradiation, of the
ionization level of the local absorbing material in the
line-of-sight. The material becomes more transparent, especially at
low energy, than it was just before the burst.  Consequently, the
subtraction of the spectrum extracted prior to the burst from the
spectrum extracted during the burst yields to an apparent soft excess
in the net spectrum. This component will not be investigated further
here. A soft excess was reported in other bursts from \src\ and
interpreted in a similar way by \cite{0748:asai06pasj}.

\begin{figure*}[!t] 
%  \centerline{\includegraphics[width=0.5\textwidth]{fig_time_kt_flux.ps}\includegraphics[width=0.5\textwidth]{fig_kt_radius_flux.ps}}
  \centerline{\includegraphics[width=0.5\textwidth]{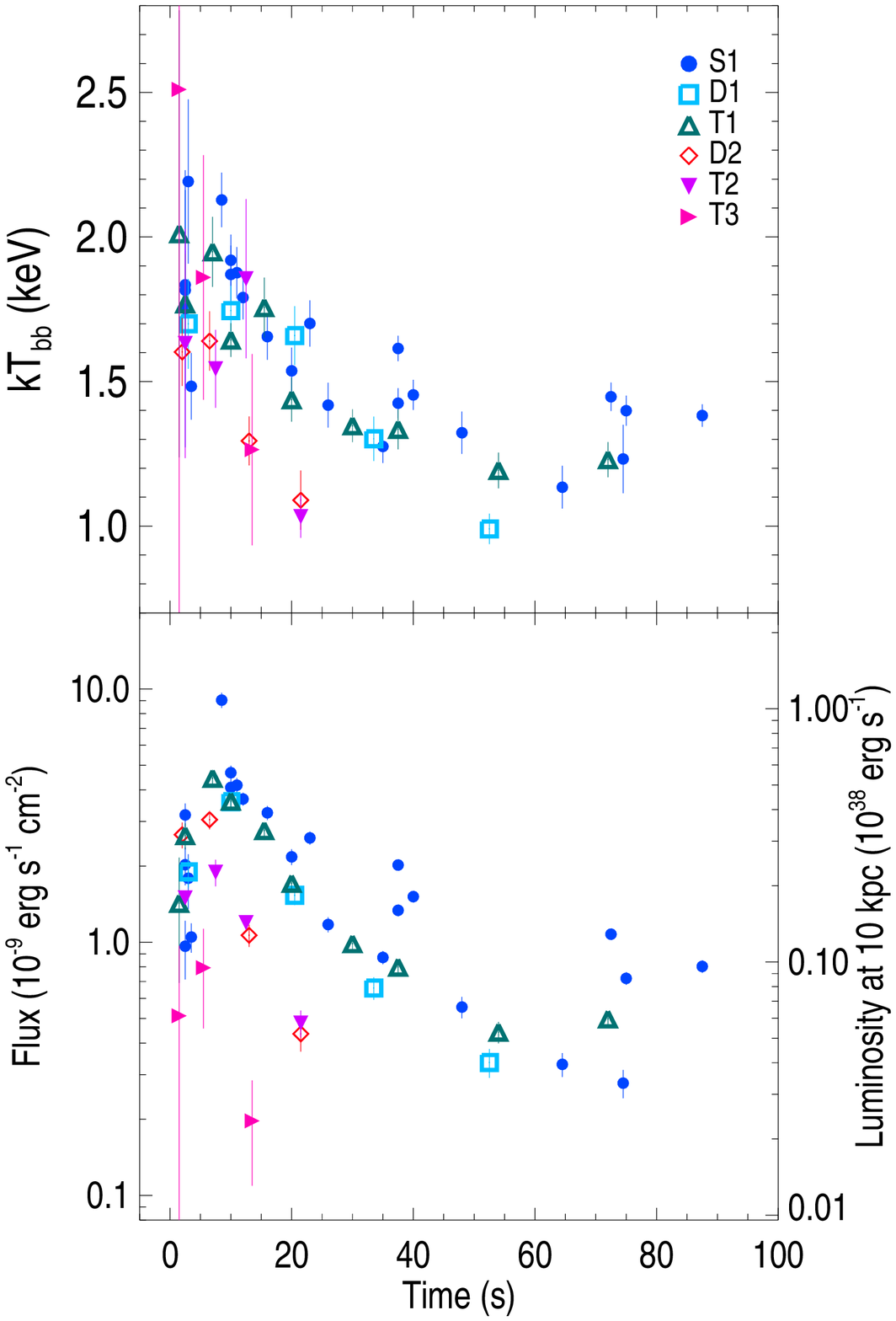}\includegraphics[width=0.5\textwidth]{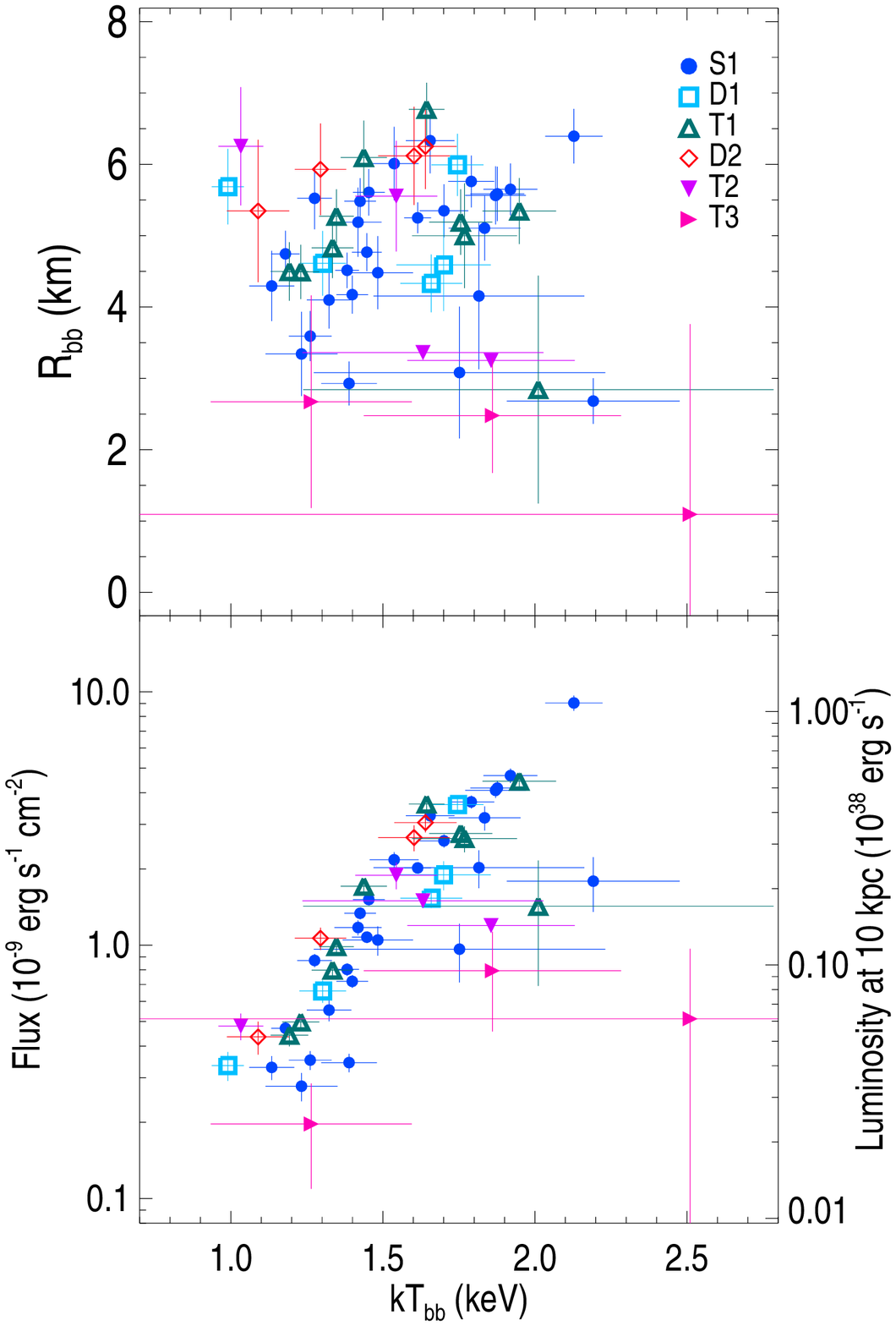}}
\caption{Results of the time-resolved spectral analysis of bursts number 1, 12,
16, 33 and 37 (S1) , 4 (D1), 5 (D2), 22 and 34 (T1), 23 (T2), and 36
(T3).  {\bfseries Left:} Blackbody temperature (top) and bolometric
flux (bottom) as a function of the time since burst onset.  {\bfseries
Right:} Blackbody radius (top) and bolometric flux (bottom) as a
function of the blackbody temperature. }
\label{fig:spectralresults}
\end{figure*}

We obtain good fits for each interval.  Fig.~\ref{fig:spectralresults}
(left) shows the derived blackbody temperatures and bolometric fluxes
as a function of time. Fig.~\ref{fig:spectralresults} (right) shows
the blackbody radius (top) and the bolometric flux (bottom) as a
function of the blackbody temperature.  For a given burst, we average
the best-fit parameter values over the intervals near the peak on one
hand, and over the intervals in the tail on the other hand.  In
Table~\ref{tab:spectralfit}, we give the range of these peak and tail
values covered by our burst sample. The blackbody surface temperature
$kT_{\mathrm{bb}}$ lies in the interval $(0.93-2.13)\pm0.11$ keV,
where higher values are found in the burst peak and lower values in
the burst tail.  There is no evidence for photospheric expansion. The
normalization of the blackbody component and hence the radius of the
emitting region is consistent with being constant throughout a burst
and consistent with being the same for all the bursts studied.  The
average radius of the blackbody component is 2.4~km assuming a source
distance of 5~kpc, and 4.9~km assuming a distance of 10~kpc, with an
average uncertainty of 44\%. This blackbody radius, since derived from
a measured color temperature rather than the effective temperature, is
an underestimate of the physical radius of the neutron star
\citep[e.g. ][]{london84apj}.

Using the previous spectral models, we further calculate the
unabsorbed bolometric flux of the blackbody component for each
interval, and finally the bolometric burst fluence by integrating over
the whole burst.  We find that, for our sample of bursts, the
bolometric fluence linearly correlates with the 5--10~keV net counts
in the burst. This means that the ratio between the bolometric fluence
and the 5--10~keV net counts in the burst is consistent with being the
same for all bursts types.  We derive a value of $(8.1\pm0.3) \times
10^{-11}$~erg~cm$^{-2}$~count$^{-1}$ for this ratio. Analogously we
find a ratio of $(8.2\pm0.3) \times
10^{-11}$~erg~cm$^{-2}$~count$^{-1}$ between the bolometric burst peak
flux and the 5--10~keV peak count rate.  These ratios serve to
estimate the bolometric fluence and peak flux from the 5--10~keV
counts for all other bursts, even when they are contaminated by
dipping.

\section{Comparison with the \exosat\ observations}
\label{sec:exosat}

In 1985, \exosat\ observed \src\ at different persistent 0.1--20~keV
fluxes between 3 and 18~$\times 10^{-10}$~\ergcms\
\citep{0748:parmar86apj,0748:gottwald86apj}.  26 bursts were observed,
including four doublets with a burst separation of the order of
10--20~min \citep{0748:gottwald86apj}. The doublet phenomenon only
occurred when the persistent 0.1--20 keV flux was below 5~$\times
10^{-10}$~\ergcms, and was part of an overall correlation of the burst
properties with changing mass accretion rate. As the persistent flux
increased, (i) the wait time of the LWT bursts increased from
$\sim$1.8 up to 16~h, (ii) their shape changed from a ``slow'' (long
tail) to a ``fast'' profile sometimes showing photospheric expansion,
(iii) their flux at a given temperature increased, and equivalently
(iv) their apparent blackbody radius at a given temperature increased.
\cite{0748:gottwald86apj} hypothesized that the persistent flux
dependent variations in the burst properties could be caused by the
flashes changing from helium-dominated at high accretion rates, to
hydrogen-triggered hydrogen-helium flashes at low accretion rates.
They further speculated that double bursts could be a feature of
hydrogen-triggered hydrogen-helium flashes.

In 1986, \exosat\ caught \src\ at a persistent 0.1--20 keV flux of
5~$\times 10^{-10}$~\ergcms\ \citep{0748:gottwald87apj}.  11 bursts
were observed and showed a regular pattern with a long recurrence time
always followed by a short one, reminiscent of the double burst
phenomenon, but with longer wait times, in the range 20--70~min for
the D2-like bursts. All the burst properties were otherwise consistent
with that of bursts in the slow mode (\ie\ low states) of 1985.

Using the spectral model by \citet{diaztrigo06aa} (see
Sect.~\ref{sec:nonbursting}), we derive an unabsorbed flux of $4.7
\times 10^{-10}$~\ergcms\ in the 0.1--20~keV band for \src\ during the
2003 \xmm\ observations, indicating that the LMXB was again in a low
state. We further suggest that the source state in 2003 is likely
closer to the lowest states of 1985 than to the low state of 1986
since the unabsorbed \exosat\ fluxes are probably underestimated
compared to the \xmm\ ones because of the local neutral and ionized
absorbers included in the \xmm\ spectral models.

The comparison of Fig.~\ref{fig:spectralresults} with Figs.~6 and 7 of
\citet{0748:gottwald86apj} and Figs.~4 and 5 of
\citet{0748:gottwald87apj} indicates that the spectral properties of
the \xmm\ bursts are all consistent with that of \exosat\ bursts in
the slow mode (low states). In particular, the \xmm\ values of the
blackbody radius ($4.9 \pm 2.1$~km on averaged) agree well with the
\exosat\ values obtained in the low states whereas larger radii of
$\sim$9~km had been obtained in higher states. These variations in the
blackbody radius for a given temperature are more likely related to a
change in the structure and composition of the neutron star outer
layers or atmosphere which distorts the blackbody spectrum than to
real variations of the emitting area \citep{0748:gottwald87apj}.  In
any case, the blackbody radius, since derived from a measured color
temperature rather than the effective temperature, is an underestimate
of the physical radius of the neutron star
\citep[e.g. ][]{london84apj}. The latter one was estimated by
\cite{0748:ozel06nat} to be $13.8\pm1.8$~km in \src.

\begin{figure}[!t]
\vspace{-0.5cm}
%\centerline{\includegraphics[angle=0,width=0.4\textwidth]{fig_histo_wt_xmm_exosat.ps}}
\centerline{\includegraphics[angle=0,width=0.4\textwidth]{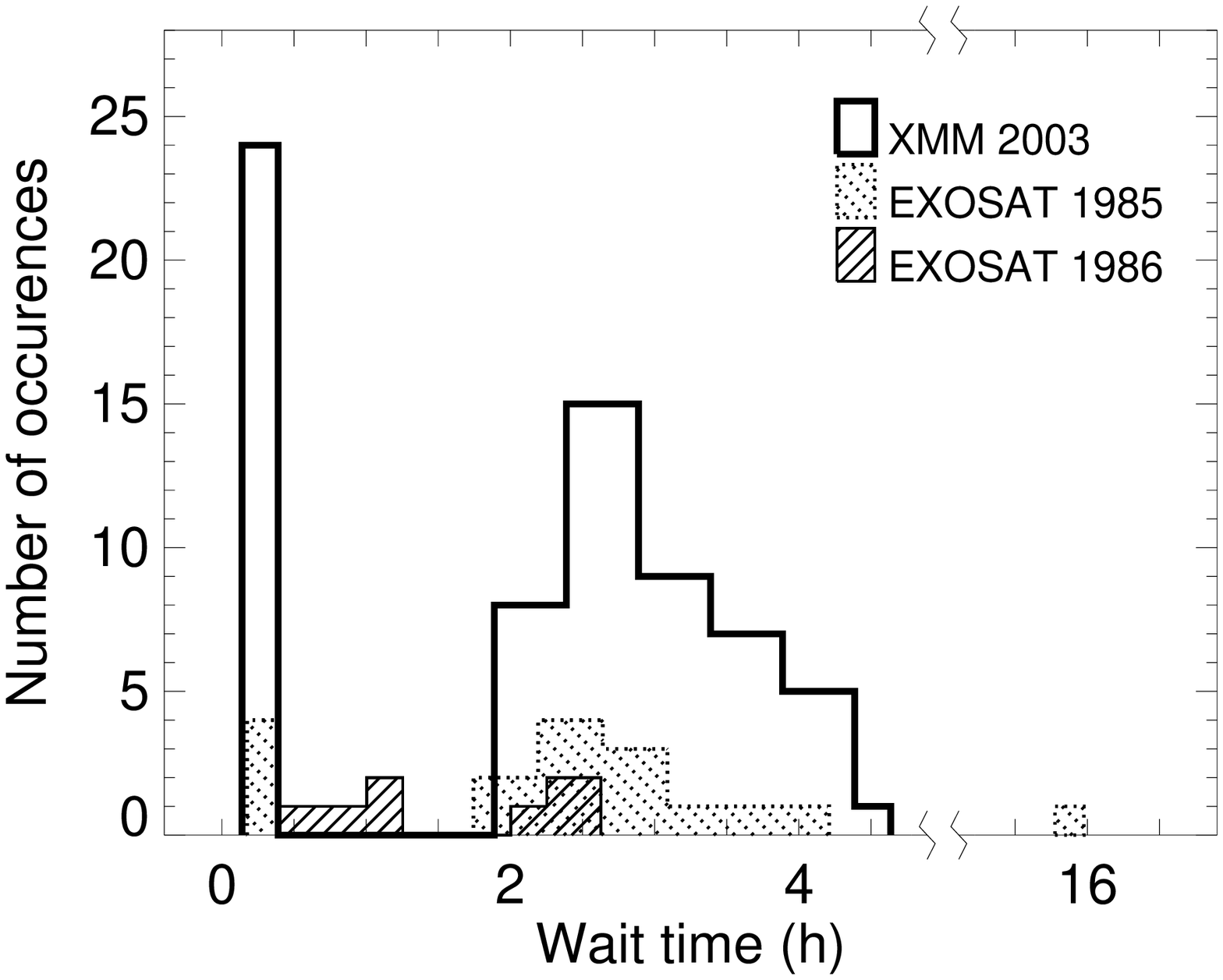}}
\vspace{-0.5cm}
%\centerline{\includegraphics[angle=0,width=0.4\textwidth]{fig_histo_wt_exosat_gap.ps}}
\centerline{\includegraphics[angle=0,width=0.4\textwidth]{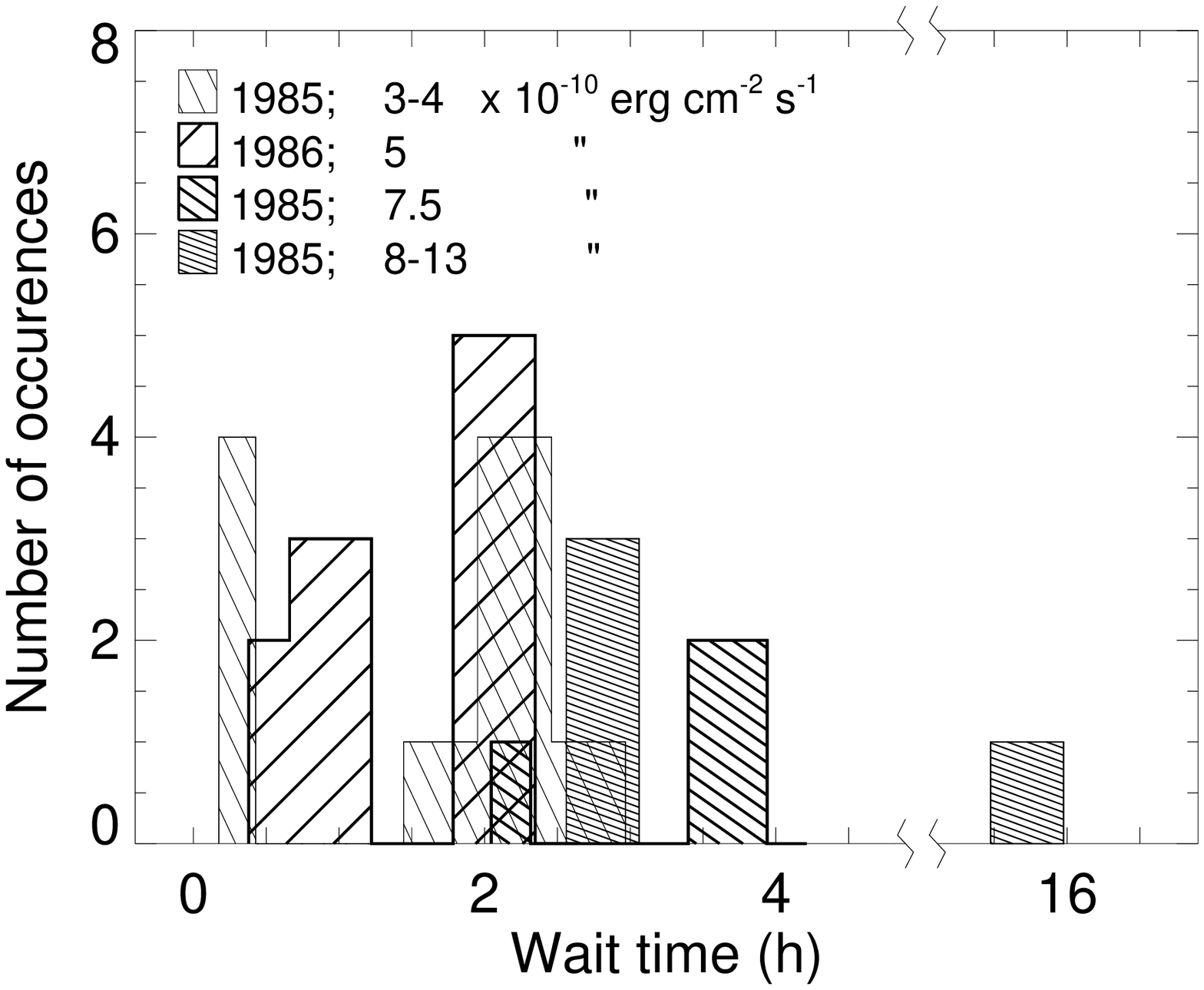}}
\caption{{\bf Top:} \src\ burst wait time distributions recorded in
1985 and 1986 by \exosat\ (from Table~2 of
\citealt{0748:gottwald86apj} and Table~1 of
\citealt{0748:gottwald87apj}) and 2003 by \xmm\ (this paper). {\bf
Bottom:} \exosat\ burst wait time distributions as a function of the
persistent 0.1--20~keV flux of \src\ indicated in the plot (from Tables~1 and 2 of
\citealt{0748:gottwald86apj}).  Note the broken wait time axis in both panels.}
\label{fig:waitexosat}
 \end{figure}

Fig.~\ref{fig:waitexosat}~(top) shows the \exosat\ burst wait time
distributions together with the \xmm\ one.  While no burst at all was
detected with a wait time between $\sim$20~min and $\sim$2~h neither
in 1985 nor in 2003, this gap was partly filled in 1986 by the D2-like
bursts with 20--70~min wait times.  Fig.~\ref{fig:waitexosat}~(bottom)
displays the \exosat\ wait time distributions as a function of the
persistent 0.1--20~keV flux.  There is a global shift of the
histograms towards higher wait times as the persistent flux increases.
The \xmm\ wait time distribution, and especially the presence of
doublets with $\sim$12~min burst separation, is consistent with that
obtained by \exosat\ when \src\ was in its lowest state in 1985.

\begin{table}[!t]
\caption{Averaged results of fits of the 0.1--10~keV
spectra extracted during intervals of 13 bursts, with a model
consisting of an absorbed blackbody and a Gaussian. The parameters are
the hydrogen column density $N_{\mathrm{H}}$, the blackbody
temperature $kT_{\mathrm{bb}}$ and normalization $N_{\mathrm{bb}}$
(defined as $R_{\rm km}^2/d_{\rm 10}^2$ where $R_{\rm km}$ is the
source radius in km and $d_{\rm 10}$ the distance to the source in
units of 10~kpc), the Gaussian centroid energy $E$, width $\sigma$ and
normalization $N_{\mathrm{G}}$. For a given burst, we averaged the
best-fit values from the intervals near the peak on one hand, and from
the intervals in the tail on the other hand. A range indicates the
spread in the obtained averaged values for the different
bursts. Uncertainties are the average errors of the parameters at
$1\sigma$ confidence level. $\chi_{\mathrm{reduced}}^{2}$ is
calculated using 50 degrees of freedom. Its average and root mean
squared are given.}
\begin{center}
\begin{tabular}{l @{\extracolsep{0.2cm}} l @{\extracolsep{0.2cm}} l}
\hline 
\hline
\noalign {\smallskip}
 & Peak & Tail \\
\hline 
\noalign {\smallskip}
$N_{\mathrm{H}}$ ($10^{22}\,\mathrm{cm^{-2}}$)&
$0.1$ (fixed) & $0.1$ (fixed)\\
$kT_{\mathrm{bb}}$ (keV)&
$(1.54-2.13)\pm0.12$ & $(0.93-1.39)\pm0.09$ \\
$N_{\mathrm{bb}}$&
$(6-49)\pm5$ & $(7-39)\pm5$ \\
$E$ (keV)&
$0$ (fixed) & $0$ (fixed) \\
$\sigma$ (keV)&
$(0.5-0.8)\pm0.6$ & $(0.40-0.84)\pm0.10$ \\
$N_{\mathrm{G}}\,(\mathrm{cts}\ \mathrm{cm^{-2}s^{-1}})$&
$(0.15-1.94)\pm0.10$ & $(0.08-0.36)\pm0.05$\\
$\chi_{\mathrm{reduced}}^{2}$ &
$1.1\pm0.2$ & $1.2\pm0.3$ \\
\noalign {\smallskip}
\hline
\end{tabular}
\end{center}
\label{tab:spectralfit}
\end{table}

These comparisons confirm the above-mentioned dependencies of the LWT
bursts properties (shape, spectral characteristics, wait times) on the
persistent flux.  The presence of SWT bursts could also depend on the
persistent flux in the following scheme.  At low persistent flux (in
the low states of 1985 and in 2003), doublets are emitted with a burst
separation of $\sim$10--20 min. Triplets occur in that regime as
well. They could have been missed by \exosat\ by chance since the
satellite looked at \src\ in the low state only 3 times $\sim$9~h,
while \xmm\ detected a triplet only once every 20~h.  In 1986, the
D2-like bursts have longer wait times of $\sim$20-70 min. This could
be related to a slightly higher persistent flux then, although the
derived fluxes in the low states of 1985, 1986 and 2003 are too close
to each other to consider this link as certain.  At much larger fluxes
(in the intermediate and high states of 1985), there are no bursts
with short wait times anymore.

\section{Discussion}
\label{sec:discu}

Our findings may be summarized as follows.  \xmm\ observed \src\ in a
low state 7 times, for an average of 22~h per observation, between
September and November 2003. 76 sub-Eddington bursts were detected,
either in single, double or triple events, respectively on 33, 14 and
5 occasions. The separation between two events is $\sim 3$~h while the
separation between two bursts in a doublet or a triplet is only
$\sim$12~min. The primary bursts of different events are similar in
terms of duration, peak flux and fluence. Their fluence is strongly
correlated with the amount of material accreted before the burst. The
profiles of these bursts show an initial fast decay and a second
slower decaying component.  Quite different are the secondary and
tertiary bursts. They have a shorter duration, lower peak flux and
lower fluence. Their fluence is not strongly correlated with the
(little) amount of material accreted before the burst, except maybe
for the tertiaries. The profiles of the secondary and tertiary bursts
show only the fast decay and lack the second, more slowly decaying,
component seen in the primaries.  There is no strong spectral
differences between the various burst types.
At a given accumulated mass, the total fluence of a triplet is
generally larger than that of a doublet which is larger than that of a
singlet.  For a new event to occur, one has to wait longer after a
doublet or a triplet than after a singlet.

\subsection{Comparison with theory: LWT bursts}

How do these results fit within the current theoretical framework? Let
us first consider the recurrence times and energetics of the LWT
bursts.  There are two ways to make bursts (or burst
``events'') with a recurrence time as short as 3 hours: mixed H/He
ignition at an accretion rate close to $10$\% Eddington, or unstable
hydrogen ignition at an accretion rate of $\approx 1$\% Eddington.

\subsubsection{Mixed H/He ignition}

 The best-studied of these is the first, unstable helium
ignition in a hydrogen-rich environment. At high accretion rates, the
accumulating fuel layer is hot enough that the hydrogen burns via the
beta-limited and thermally stable hot CNO cycle
\citep{hoyle65conf}. When the helium ignites, it burns in a hydrogen
rich environment, prolonging the nuclear energy release and leads to
long burst tails.

The shape of the burst lightcurves from \src\ is similar to those
observed from \gs\ which \cite{1826:galloway04apj} argued is bursting
in this regime.  The observed $\alpha$ value supports this
interpretation, since the typical energy release $Q_{\rm nuc}\approx
4$ MeV per nucleon in the rp-process (\eg, \citealt{schatz99apj};
Table~10 of \citealt{woosley04apj}) gives an expected $\alpha$
\cite[\eg,][]{1826:galloway04apj}) close to the observed value
$\alpha=53$:
\begin{equation}\label{eq:alpha}
\alpha={c^2z\over Q_{\rm nuc}}\left({\xi_b\over\xi_p}\right)=48\
\left({z\over 0.31}\right)\left({4\ {\rm MeV}\over Q_{\rm
nuc}}\right)\left({\xi_p/\xi_b\over 1.5}\right)^{-1},
\end{equation}
where $z$ is the gravitational redshift, and $\xi_{b,p}$ are factors
to account for anisotropy in the burst and persistent
luminosities. They are defined by $L_{b,p} = 4\pi d^2 \xi_{b,p}
f_{b,p}$, where $L_{b,p}$ and $f_{b,p}$ are the intrinsic luminosities
and the observed fluxes, and $d$ the distance to the source
\citep[][]{lapidus85mnras,fujimoto88apj}\footnote{\label{noteozel} We
note that anisotropy of the burst luminosity was not included in the
analysis by \citet{0748:ozel06nat}. The \src\ neutron star mass and
radius determinations in that paper are likely unaffected by the
anisotropy factor, as long as it is constant through the burst, since
they come from taking the ratio of the peak flux and the flux in the
tail. However, the lower limit on the source distance, $d>9.2\pm 1.0\
{\rm kpc}$ should be reduced by a factor $\xi_b^{1/2}$, or 15--30\%,
giving $d>7.3$--$8.1\ {\rm kpc}$ for $\xi_b=1.3$--$1.6$, using the
estimate of \cite{fujimoto88apj}.}.  $\xi_{b,p}>1$ indicates that the
observed flux is lower than it would be in the absence of anisotropy.
Here we use the estimates $\xi_b^{-1}=0.5+\left|\cos i\right|$ and
$\xi_p^{-1}=2\left|\cos i\right|$, obtained by \cite{fujimoto88apj}
for a system viewed under inclination $i$ whose geometry enhances the
emission in the direction perpendicular to the disk plane, because of
scattering in the disk, more strongly when the incident photons come
from the inner disk (during persistent emission) than when they come
from the neutron star surface (during bursts). Using $i\approx
75$--$83^\circ$ for \src\ \citep{0748:parmar86apj} gives $\xi_b\approx
1.3$--$1.6$, $\xi_p\approx 1.9$--$4.2$, and a ratio
$\xi_p/\xi_b\approx 1.5$--$2.6$. We note that the $\xi_{b,p}$
estimates could significantly differ if other system geometries (\eg\
including disk warping) were assumed.

In multizone X-ray burst simulations of mixed H/He ignition bursts, \cite{woosley04apj} found
recurrence times close to 3 hours for accretion of solar metallicity
material at $\dot M=1.75\times 10^{-9}\ M_\odot\ {\rm yr^{-1}}$, or
luminosity $\dot M(GM/R)\approx 2\times 10^{37}\ {\rm erg\ s^{-1}}$
for a $1.4\ M_\odot$, 10 km neutron star (models zM and ZM of
\citealt{woosley04apj}). This agrees quite well with the observed
X-ray luminosity from \src, once the anisotropy parameter is
included. The unabsorbed $0.1$--$100\ {\rm keV}$ flux is $8.4\times
10^{-10}\ {\rm erg\ cm^{-2}\ s^{-1}}$, giving a luminosity
$L_X=0.48\times 10^{37}\ {\rm erg\ s^{-1}}\ (\xi_p/1.9)$ for a
distance of $5\ {\rm kpc}$ and $L_X=1.9\times 10^{37}\ {\rm erg\
s^{-1}}\ (\xi_p/1.9)$ for a distance of $10\ {\rm kpc}$, where we
normalize the anisotropy parameter to $i=75^\circ$.
Therefore we find that the recurrence times and $\alpha$ values of the
LWT bursts agree well with the models of \cite{woosley04apj},
particularly when the anisotropy factor is included.

However, the total energies of the LWT bursts in \src\ are somewhat
lower than in the models, or than observed for \gs. In the mixed H/He
regime, the ignition column depth is $y_{\rm ign}\approx 2\times 10^8\
{\rm g\ cm^{-2}}$ \citep{cumming00apj} giving a burst energy $E_{\rm
burst}=4\pi R^2Q_{\rm nuc}y_{\rm ign}/(1+z)$ $=7.4\times 10^{39}\ {\rm
ergs}$ (assuming complete burning of the whole surface, radius $R=10\
{\rm km}$, and energy release $Q_{\rm nuc}=4\ {\rm MeV}$). This
estimate agrees well with models zM and ZM of \cite{woosley04apj} (see
their Table 9) and with the observed burst energies in \gs\
($5.3\times 10^{39}\ {\rm ergs}$ at $6\ {\rm kpc}$,
\citealt{1826:galloway04apj}), while the mean burst energy for the LWT
bursts in \src\ is $(0.5$--$2.1)\times 10^{39}\ {\rm ergs}$ for the
distance range $5$--$10\ {\rm kpc}$.  Note that anisotropy, which
would tend to decrease the burst luminosity for \gs\
\citep{1826:galloway04apj} and increase it for \src\
\citep[][]{fujimoto88apj}, is not included in these estimates.

The same is true for the peak luminosity. The high base temperatures
$\gtrsim 10^9\ {\rm K}$ reached in mixed H/He bursts give a peak
luminosity $\gtrsim 10^{38}\ {\rm erg\ s^{-1}}$
\citep[\eg,][]{woosley04apj}, approaching the Eddington
luminosity. Ignoring anisotropy, in \gs, the bursts peak flux was
$3\times 10^{-8}\ {\rm erg\ cm^{-2}\ s^{-1}}$, giving $L_{\rm
peak}=1.3\times 10^{38}\ {\rm erg\ s^{-1}}$ at $6\ {\rm kpc}$
\citep[][]{1826:galloway04apj}. In contrast the mean peak flux for the
LWT bursts in \src\ is  7.2$ \times 10^{-9}\ {\rm ergs\
cm^{-2}\ s^{-1}}$, giving $L_{\rm peak}=(0.22$--$0.86)\times
10^{38}\ {\rm erg\ s^{-1}}$ for the distance range $5$--$10\ {\rm
kpc}$. Independent of the assumed distance, it is a factor of 7 below
the peak flux of the bright radius expansion burst observed by
\cite{0748:wolff05apj} from \src, further showing that these LWT
bursts are faint relative to the Eddington luminosity.

\subsubsection{Hydrogen ignition}

The low burst energies and peak luminosities suggest a different
explanation for the LWT bursts from \src, that they are triggered by
unstable hydrogen ignition. Hydrogen burning is unstable
for temperatures less than $\approx 8\times 10^7\ {\rm K}$, when the
CNO is no longer beta-limited, and can trigger thermonuclear flashes
\citep{fujimoto81apj}.

\cite{bildsten98conf} estimates that
unstable hydrogen ignition occurs for accretion rates $\dot M\lesssim
2\times 10^{-10}\ M_\odot\ {\rm yr^{-1}}$, or luminosities
$L_X\lesssim 2\times 10^{36}\ {\rm erg\ s^{-1}}$. This matches the
luminosity of \src\ if the source is located at the closer end of its
distance range, and the anisotropy factor is small. The observed
recurrence time of $\approx 3\ {\rm hours}$ matches the recurrence
time expected close to the transition between unstable and stable
hydrogen burning. The maximum temperature at which hydrogen can
unstably ignite, close to $8\times 10^7\ {\rm K}$, corresponds to a
column depth of approximately $10^7\ {\rm g\ cm^{-2}}$ (see Fig.~1 of
\citealt{cumming04conf}). At an accretion rate of 1\% of the Eddington
rate, this column is accreted in $3.2$ hours.

There are two possible outcomes of unstable hydrogen ignition
\citep{fujimoto81apj,peng06apj}. If the hydrogen ignition depth is
$\gtrsim 5\times 10^7\ {\rm g\ cm^{-2}}$, the minimum column depth at
which helium can ignite (Fig.~1 of \citealt{cumming04conf}), the
increase in temperature following hydrogen ignition triggers ignition
of helium by the triple alpha reaction, and a mixed hydrogen/helium
flash occurs. At smaller ignition depths, which occur near the
transition between unstable and stable hydrogen burning, the hydrogen
flash is not able to trigger helium ignition. This case has recently
been modeled by \cite{peng06apj}. They include, for the first time,
sedimentation of heavy elements in the accumulating fuel layer. For an
accretion rate close to $1$\% of the Eddington rate, they find that
the hydrogen flash reaches a peak luminosity of 5 times the accretion
luminosity, or $\approx 0.1\times 10^{38}\ {\rm erg\ s^{-1}}$. This is
lower than the observed peak luminosities of the LWT bursts by a
factor of a few. Burning a column depth of $10^7\ {\rm g\ cm^{-2}}$ of
hydrogen to helium gives an expected burst energy $0.6\times 10^{39}\
{\rm ergs}$ (using the energy release $Q_{\rm nuc}=6.0\times 10^{18}\
{\rm erg\ g^{-1}}$ appropriate for the hot CNO cycle). This is within
the range of observed energy for the LWT bursts.

Although the peak luminosity found by \cite{peng06apj} is a little
lower than observed for the LWT bursts, they only computed a few
different cases of hydrogen ignition. In addition, the recurrence time
and burst energies do match the expectations for hydrogen ignition
close to the stability boundary quite well. An interesting point about
this interpretation is that the sedimentation of heavy elements plays
a crucial role \citep{peng06apj}. Without sedimentation, the CNO
abundance remains close to the solar value at the ignition depth, and
the energy released by the initial proton captures is small giving a
very weak flash. With sedimentation, the CNO abundance is enhanced by
a factor of several, leading to more energy release during the initial
runaway, giving a peak temperature of $\approx 3\times 10^8\ {\rm K}$
and a peak luminosity observable above the accretion luminosity.

In addition, the interpretation of the LWT bursts as hydrogen flashes
also provides two explanations for the bright energetic burst ($E_{\rm
b} = 3.6\times 10^{-7}$~\ergcm) reported by \cite{0748:wolff05apj}
from \src\ at a low persistent flux ($f_{\rm p} = 4.8 \times
10^{-10}$~\ergcms). First, if the hydrogen flash fails to ignite
helium, a pure helium layer builds up beneath the hydrogen burning
shell which will eventually ignite by triple alpha
reactions. Alternatively, since the burst properties indicate that the
hydrogen ignition is occurring close to the transition between
unstable and stable hydrogen burning, a slight increase in accretion
rate would lead to stable hydrogen burning and a similar accumulation
of a pure helium layer. Second, if the accretion rate drops, the
hydrogen ignition column depth becomes large enough to trigger helium
ignition resulting in a mixed H/He flash.  \cite{0748:wolff05apj}
derive a burst fluence of $3.6\times 10^{-7}\ {\rm erg\ cm^{-2}}$
which translates to a burst energy of $(1.5$--$2.5)\times 10^{39}\
{\rm ergs}$ (depending on whether the peak luminosity corresponds to
the Eddington limit for solar composition or for pure helium; the
corresponding distances are 5.9 and $7.7\ {\rm kpc}$). Assuming
$Q_{\rm nuc}=1.6$ MeV, appropriate for helium burning to iron group,
the implied thickness of the fuel layer is $\approx 10^8\ {\rm g\
cm^{-2}}$. A pure helium layer accumulating beneath the hydrogen shell
would typically reach much greater thicknesses before
igniting. Therefore, it seems likely that the \cite{0748:wolff05apj}
burst is another example of a hydrogen ignition, but this time also
igniting the helium.

\subsection{Comparison with theory: SWT bursts}

We have spent some time discussing the origins of the LWT bursts in
the hope that this might give a clue to the origin of the SWT
bursts. These have so far evaded theoretical explanation, although
possibilities have been put forward in the
literature. \cite{1636:fujimoto87apj} suggested that unburned fuel
left over from the preceding burst is mixed downwards by hydrodynamic
instabilities driven by rotational shear. \cite{wallace84} suggested
that fresh fuel could be mixed downwards by Rayleigh-Taylor
instabilities which set in as the layer cools. However, neither of
these explanations offers a natural explanation for the ten minute
delay time. This point is emphasized dramatically by the discovery of
burst triplets discussed in this paper. The fact that the same
characteristic timescale sets the delay between the second and third
bursts in a triplet as well as the first and second points to some
physical process with this timescale that can recur more than once in
succession.

Mixed H/He bursts at accretion rates close to 0.1 of the Eddington
rate have been modeled extensively in spherical symmetry with large
nuclear reaction networks (Woosley et al.~2004). However, there has
been little work on hydrogen triggered bursts, and so the suggestion
that the LWT bursts are hydrogen triggered opens up the possibility
that the ten minute phenomenon is connected with unstable hydrogen
burning. For example, the half-life of $^{13}$N in the CNO cycle is
9.97 minutes, very close to the observed timescale. Perhaps as the
layer cools following the initial flash, an instability driven by
proton captures on $^{13}$C or $^{14}$N occurs once there has been
time for enough seed nuclei to be produced by the beta-decay of
$^{13}$N.  However, it is not clear how the fact that the timescale
between SWT bursts changes (see Sect.~\ref{sec:exosat}), possibly with
accretion rate, would be accommodated in this picture. Further
theoretical studies of hydrogen triggered bursts are needed to explore
the possibility that the SWT bursting has a nuclear physics
explanation.

Even without a good picture of the physics of the SWT bursts, we can
explore some possibilities by considering the idea that the second and
third bursts are caused by ignition of leftover fuel from the first
burst.  One way that this might happen is that the ignition conditions
for the first burst are always the same, but that sometimes incomplete
burning occurs, leaving behind unburned fuel that later reignites
leading to double or triple events.  However, this is inconsistent
with the observation that, at a given wait time, the fluence is the
same for a D1, a T1 or a S1 burst (the LWT bursts have all the same
alpha values).  In addition, the fact that the wait time to the next
event is longer after a double or triple (regardless of the type of
the next event) also suggests a different picture.  Imagine that after
the first burst (either S1, D1, or T1), a fraction $f$ of the fuel
layer is left unburned. Then, in double or triple events, this
leftover fuel layer is burned after a delay time of $\approx 10$
minutes, whereas in single events, the unburned fuel survives, and
leads to early ignition of the next event.  The difference between
singles and doubles/triples here is not whether there is any unburned
fuel, but instead whether the unburned fuel is able to ignite on the
ten minute timescale.

One way to test this idea is to look for consistency between the wait
times and the burst fluences. The wait time after a double (triple) is
$\approx 37$\% (55\%) longer than the wait time after a single,
suggesting that a fraction $f_a \approx 37$\% (55\%) of the available
accreted fuel is not burned during the first burst.  If this unburned
fuel was of the same composition as the rest of the layer and burned
subsequently in the second/third bursts, we would expect the
double/triple events to have a larger fluence by the same factor
$f_a$. Now, the double (triple) events have larger fluences than the
singles, but only by a factor $f_b$ of $\approx 17$\% (25\%).  This
indicates that the layer that burns in the second/third bursts has a
lower energy per gram, for example a smaller hydrogen content than the
accreted fuel that burns in the first burst. This suggests that in
fact the ``unburned'' fuel layer undergoes some hydrogen burning into
helium during the first burst.  Interestingly, the ratio $f_a / f_b$
is $\approx 2.2$ for both the doubles and the triples and is $\lesssim
2.5$ which is the value of the ratio between the energy released by an
hydrogen nucleon burning to the iron group (4~MeV) and the energy
released by an helium nucleon burning to the iron group (1.6~MeV).
The burst lightcurves also support the idea that the second/third
bursts have a smaller hydrogen fraction since they have a much shorter
tail than the first bursts.

These properties would be naturally explained if the first burst in
the sequence involved unstable hydrogen burning into helium.  If
hydrogen ignition is occurring near the transition from unstable to
stable hydrogen burning, then it is not clear whether helium is
burning at all in the first flash. Certainly, it is not able to ignite
unstably at low column depths but maybe it burns slightly at the peak
temperature.  The first flash would thus leave a large part (if not
all) of the available helium unburned and further produce some.  A
subsequent second/third burst would take place in an helium-rich
environment. The fact that little material can be accreted within ten
minutes suggests that the SWT bursts are also hydrogen triggered since
lower column depths are required than for helium ignition.  More
theoretical work on hydrogen ignition bursts is needed to test these
ideas.

One important question concerns the occurrence rate of
quadruple events. We did not observe a quadruple burst from
\src. \cite{galloway06apjs} report one quadruple event from
4U~1636--536 in the RXTE burst catalog. The occurrence rate of
quadruple versus triple events (as well as triple versus double
events) could provide an important constraint on theoretical models.

\section*{Conclusion}

This is the first time that unambiguously and repeatedly triple bursts
are detected in an accreting neutron star, \src.  The source was
already known to frequently exhibit double bursts, where two bursts
are separated by only 12 minutes. Now the same time scale is seen in
triple bursts. This recurrence time is too short to accrete a
sufficient amount of fuel, which indicates that there must have been
fuel left unburned in the previous burst.  We suggest that after each
burst there is an amount of fuel left unburned, which may or may not
reignite after 12 minutes, in the former case producing a double or
triple burst. We find evidence that this unburned fuel yields less
energy per gram, pointing to a lower hydrogen content than for the
fuel of the primary bursts.

The recurrence times of 3 hours and the low alpha values are
consistent with mixed hydrogen/helium burning at 0.1 Eddington, as
seen for example in \gs. The persistent luminosity of \src\ is
consistent with this accretion rate if the X-ray emission from the
system is anisotropic, as predicted for large inclination angles.
However, the expected energies and peak luminosities are somewhat larger than
observed in \src. In addition, this burning regime has been well-studied theoretically and
does not predict the ten minute bursting phenomenon. New physics, such as mixing of
fuel to deeper layers, is required to explain the multiple events
within the mixed hydrogen/helium burning regime.

Hydrogen ignition at a lower accretion rate may provide a more natural explanation for the burst behavior of \src. Hydrogen ignition bursts are not
well-studied, but the properties of the bursts from \src\ agree quite
well with the recent work by \cite{peng06apj}. In particular, the low peak luminosity and burst energies of the LWT bursts, and the low persistent luminosity are consistent with burning in this regime.  This opens up the
exciting possibility that the ten minute recurrence time bursts are in
fact a natural consequence of hydrogen ignition bursts.

%------------------------------------
%       Acknowledgements
%------------------------------------

\begin{acknowledgements}
  This work is based on observations obtained with XMM-Newton, an ESA
  science mission with instruments and contributions directly funded
  by ESA member states and the USA (NASA). SRON is supported
  financially by the NWO, the Netherlands Organization for Scientific
  Research.  AC would like to thank Brandon Helfield for help with
  understanding hydrogen ignitions, and Ed Brown for sharing results
  on hydrogen triggered bursts prior to publication. AC is an Alfred
  P.~Sloan Research Fellow, and is grateful for support from an NSERC
  Discovery Grant, Le Fonds Qu\'eb\'ecois de la Recherche sur la
  Nature et les Technologies, and the Canadian Institute for Advanced
  Research.
\end{acknowledgements}

%------------------------------------
%       References
%------------------------------------

\bibliographystyle{aa}
%\bibliography{mybib}

 %
 % Appendix
 % 

 \begin{appendix}

 \section{Light curves \& bursts lists}

 Fig.~\ref{afig:rev_lc} shows the EPIC PN light curves of \src\ with a
 binning of 60~s during \xmm\ revolutions 692 to 719. Panels~a~and~b
 show the 5--10~keV and the 0.3--5~keV energy band,
 respectively. Panel~c shows the hardness ratio.  Times are not
 barycentre-corrected.  A cycle number tracing the eclipse timing is
 indicated on the top axis.  This was determined from a reference time
 of 13.79~h on 2003 September 19 (or \xmm\ time of $1.8036643e+08$~s)
 estimated as the mid-time of the first eclipse observed during
 revolution 692, and using an orbital period of 3.824~hr
 \citep{0748:wolff02apj}.  Integer values of the cycle number
 correspond to phase 0, or estimated eclipse mid-times.

 \begin{figure*}[!t]
 %\centerline{\includegraphics[angle=90,width=1.\textwidth]{fig_orbit0692_v2.ps}}
 \centerline{\includegraphics[angle=90,width=1.\textwidth]{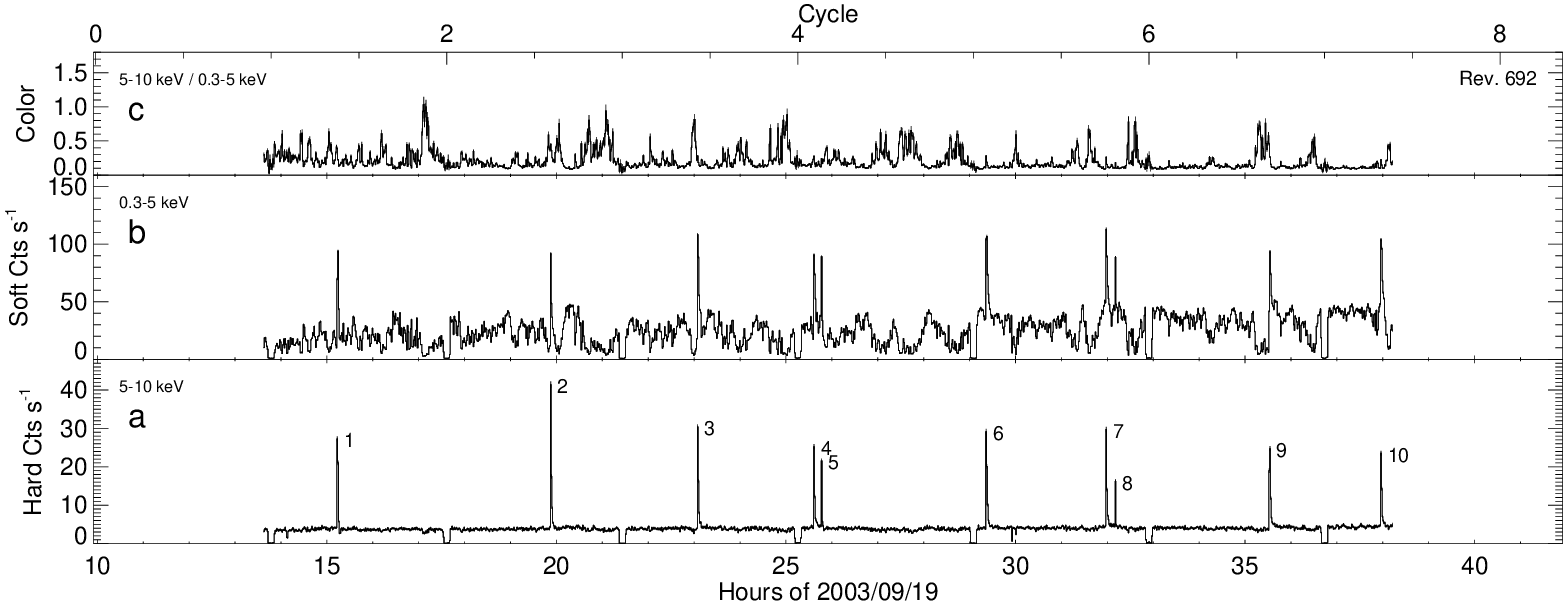}}
 %\centerline{\includegraphics[angle=90,width=1.\textwidth]{fig_orbit0693_v2.ps}}
 \centerline{\includegraphics[angle=90,width=1.\textwidth]{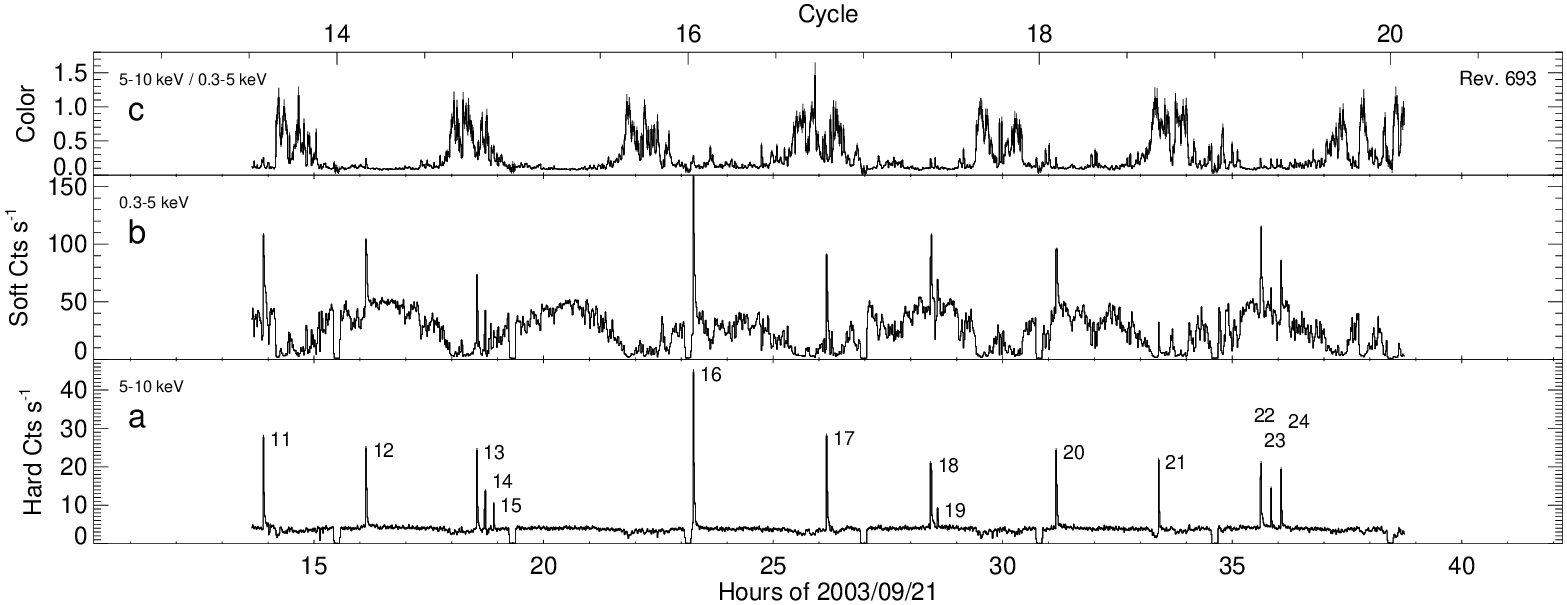}}
 \caption{\src\ as observed by \xmm\ EPIC PN during revolutions
 692 ({\it top}) and 693 ({\it bottom}). {\bfseries a)} 5--10~keV light
 curve. The bursts are numbered. {\bfseries b)} 0.3--5~keV light
 curve. {\bfseries c)} Color (counts in the 5--10~keV band divided by
 counts in the 0.3--5~keV band) as a function of time. The binning time
 is 60~s in each panel. The cycle number is indicated on the top axis,
 with the estimated mid-eclipse time of the first eclipse observed
 during revolution 692 taken as cycle number 1.}
 \label{afig:rev_lc}
 \end{figure*}

 \addtocounter{figure}{-1}
 \begin{figure*}[!t]
 %\centerline{\includegraphics[angle=90,width=1.\textwidth]{fig_orbit0694_v2.ps}}
 \centerline{\includegraphics[angle=90,width=1.\textwidth]{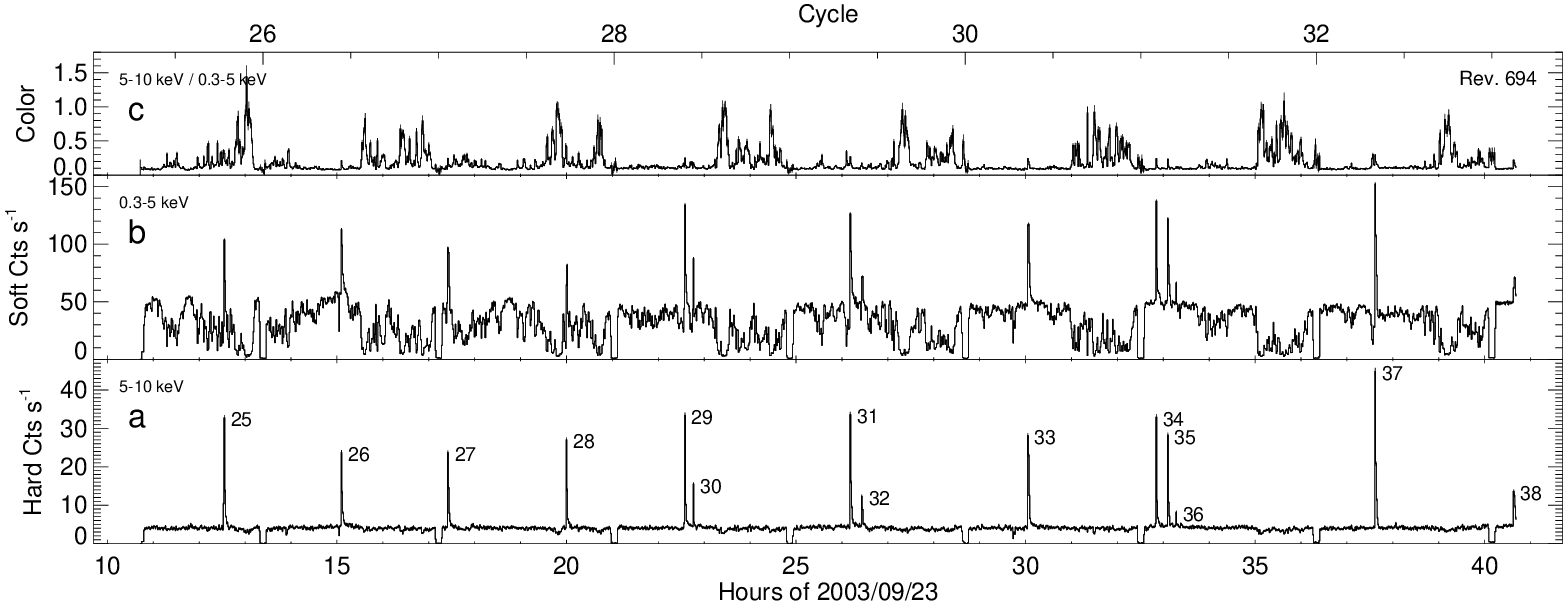}}
 %\centerline{\includegraphics[angle=90,width=1.\textwidth]{fig_orbit0695_v2.ps}}
 \centerline{\includegraphics[angle=90,width=1.\textwidth]{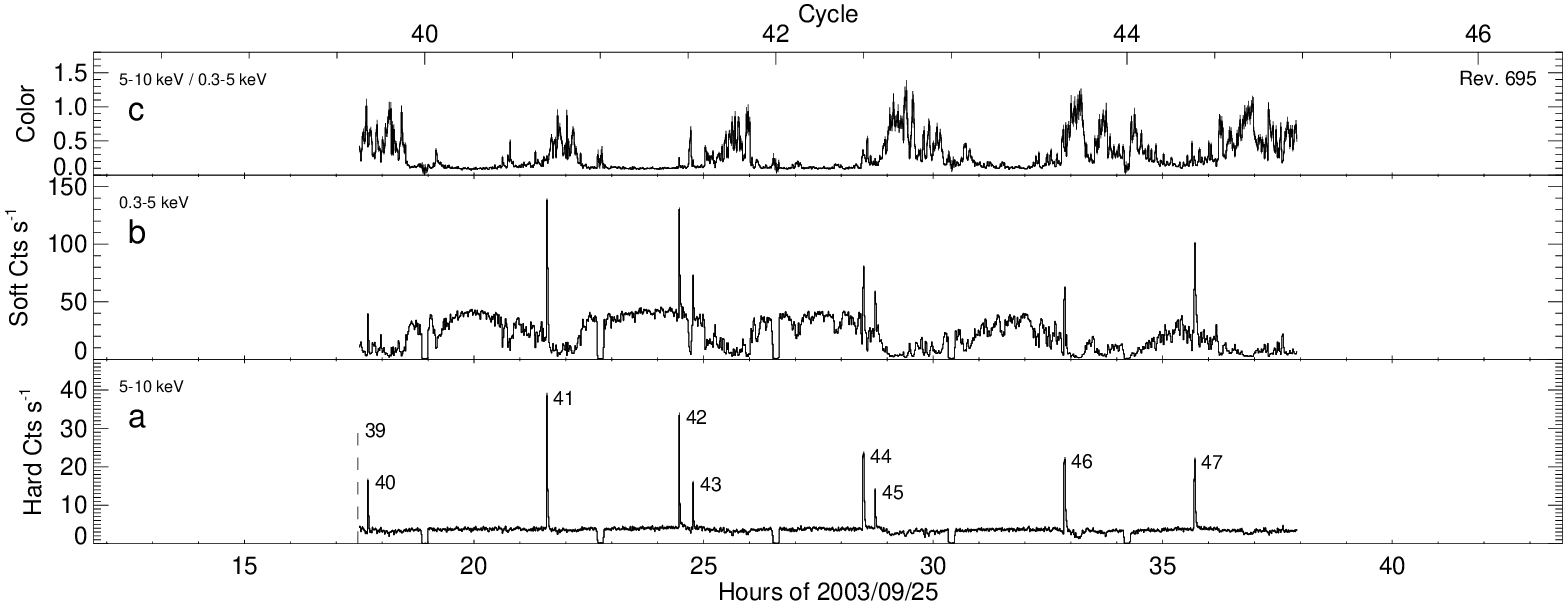}}
 \caption{Continued for \xmm\ revolutions 694
   ({\it top}) and 695 ({\it bottom}).  The dashed line in panel c of
   the bottom plot indicates the time of a burst (numbered 39) detected
   by RGS just before the start of the EPIC PN observation.}
 \label{afig:rev_lc}
 \end{figure*}

 \addtocounter{figure}{-1}
 \begin{figure*}
 %\centerline{\includegraphics[angle=90,width=1.\textwidth]{fig_orbit0708_v2.ps}}
 \centerline{\includegraphics[angle=90,width=1.\textwidth]{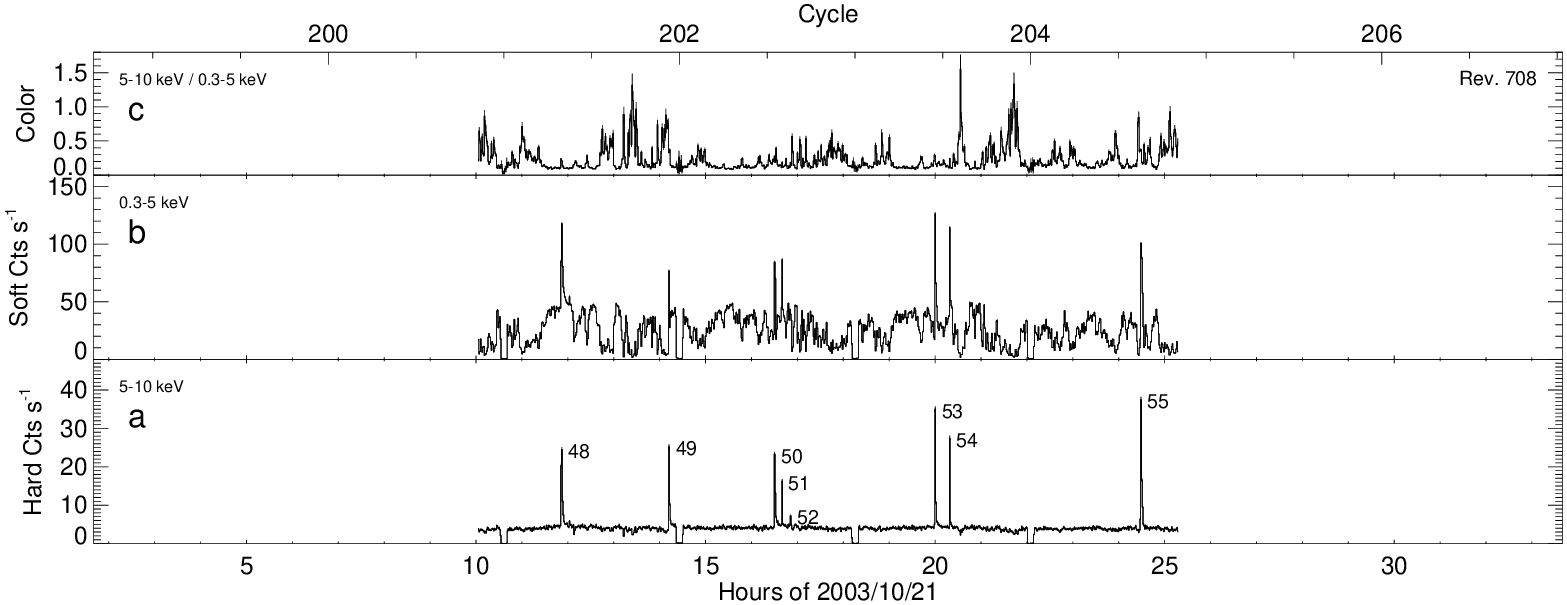}}
 %\centerline{\includegraphics[angle=90,width=1.\textwidth]{fig_orbit0710_v2.ps}}
 \centerline{\includegraphics[angle=90,width=1.\textwidth]{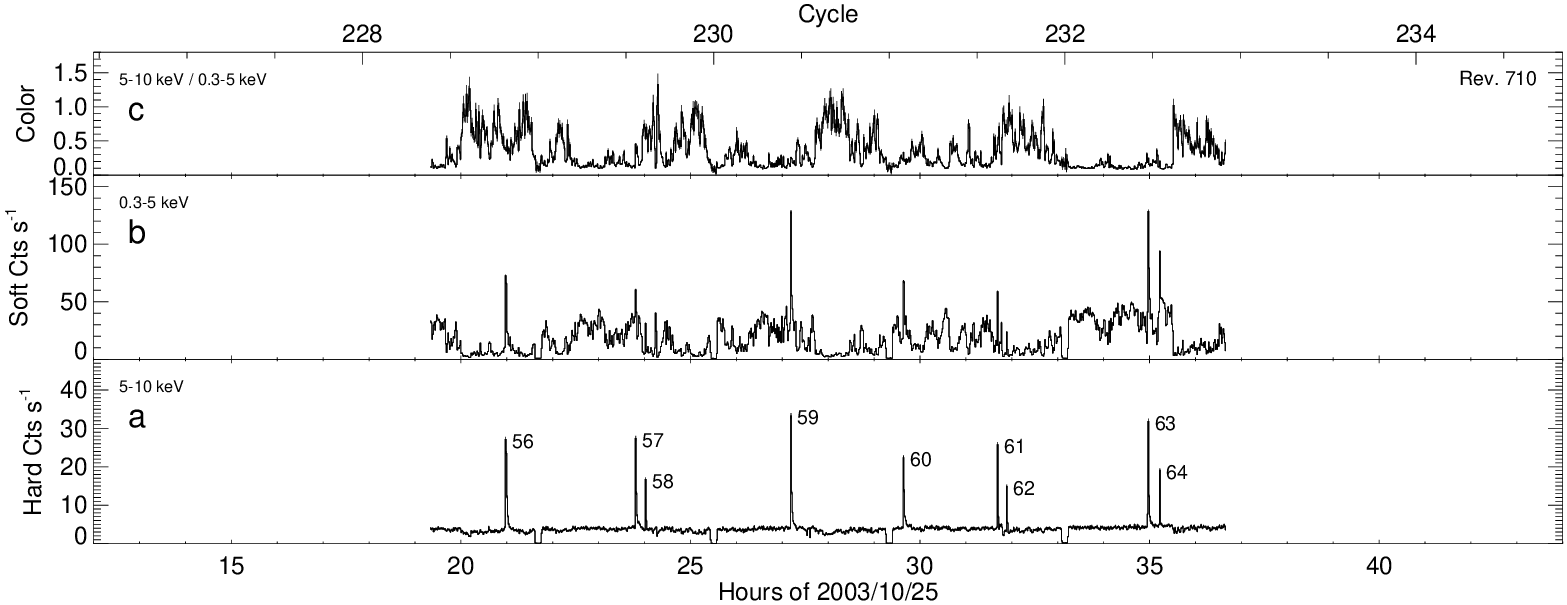}}
 %\centerline{\includegraphics[angle=90,width=1.\textwidth]{fig_orbit0719_v2.ps}}
 \centerline{\includegraphics[angle=90,width=1.\textwidth]{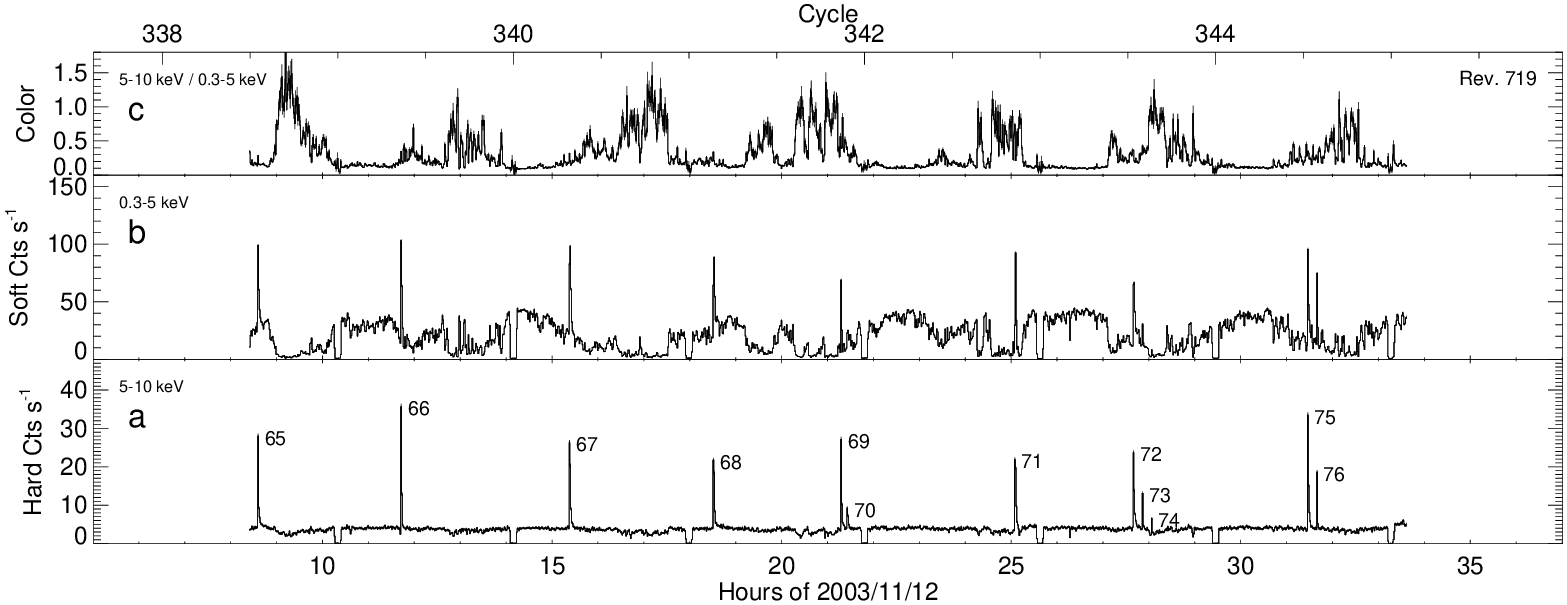}}
 \caption{Continued  for \xmm\ revolutions 708 ({\it top}), 710 ({\it middle}) 
 and 719 ({\it bottom}).}
 \end{figure*}

 The peak time, wait time and type of all the bursts are listed in
 Table~\ref{atab:times}. Since the wait time of the first burst of each
 observation in the series cannot be determined, a lower limit is given
 as the separation between the burst peak time and the start time of
 the observation.  An upper limit is given as the separation between
 the burst peak time and peak time of the last burst observed during
 the previous observation of the series.

 We note that because of scattering, residual emission is present
 during eclipses at a level of $\sim$4\% of the source's persistent
 flux.  This explains why an X-ray burst was once detected during an
 eclipse with \exosat\ \citep{0748:gottwald86apj}.  This implies that
 any S1, D1 or T1 (ie. strong) burst occurring during an eclipse of the
 \xmm\ observation would have been detected.  Since we do not detect
 any burst during eclipses, it appears unlikely that we have missed an
 S1, D1 or T1 burst because of eclipses.  Furthermore, since an eclipse
 lasts $\sim$500~s, which is less than the typical separation of the
 bursts in a multiple burst event, it is not possible either to have
 missed a complete doublet or triplet. From the inspection of
 Fig.~\ref{afig:rev_lc}, we can therefore conclude that it is unlikely
 that any S1, D1, T1, D2, T2 or T3 burst has been missed because of an
 eclipse.

 However, bursts occurring close to (less than 25~min) the beginning or
 the end of an observation, or close to a short instrumental gap such
 as occasionnally present in the data, cannot have their type
 determined certainly because another burst of the same event could
 have occurred during the gap and been missed.  In such cases, the
 different possible types are indicated in Table~\ref{atab:times}
 starting with the one attributed to the burst in this analysis.

 \renewcommand{\baselinestretch}{0.5}
 \begin{table*}[!h]
 \caption{Bursts timing. The columns indicate
 the burst number, the \xmm\ revolution number at which the burst
 occurred, the burst peak time determined from the EPIC PN 5--10~keV 1~s
 resolution light curve (except for burst 39 detected only by RGS), the
 wait time and the burst type. S means single burst. D1 and D2 mean
 first and second burst in a doublet. T1, T2 and T3 mean first, second
 and third burst in a triplet, respectively. When the burst type cannot
 be determined certainly, we indicate the different possible types,
 starting with the one attributed to the burst in this analysis.}
 \begin{center}
 \begin{tabular}{rl@{\extracolsep{0.1cm}}r@{\extracolsep{0.1cm}}r@{\extracolsep{0.0cm}}rr@{\extracolsep{0.0cm}}r@{\extracolsep{0.0cm}}ll@{\extracolsep{0.2cm}}ll}
 \hline
 \hline
 \noalign {\smallskip}
 Burst & Rev. & \multicolumn{3}{c}{Peak time (UT)} & & $t_{\rm wait}$ (h)& & \multicolumn{2}{l}{Type} & \\
 \noalign {\smallskip}
       1 &  692  &  Sep  &  19  &  15:13:58  &  1.62 $<$  &&  & S1 &  & \\
       2 &    &    &    &  19:52:12  & & 4.64 && S1 &  & \\
       3 &    &    &    &  23:04:31  & & 3.21 && S1 &  & \\
       4 &    &    &  20  &  01:36:43  & & 2.54 && D1  & (T2)	&  \\
       5 &    &    &    &  01:46:28  & & 0.162 && D2  & (T3)	&  \\
       6 &    &    &    &  05:21:42  & & 3.59 && S1 & (D2)	&  \\
       7 &    &    &    &  07:58:35  & & 2.61 && D1  &  & \\
       8 &    &    &    &  08:10:17  & & 0.195 && D2  &  & \\
       9 &    &    &    &  11:31:58  & & 3.36 && S1 &  & \\
      10 &    &    &    &  13:57:49  & & 2.43 && S1 & (D1, T1)	&  \\
      11 &  693  &    &  21  &  13:53:56  &  0.259 $<$  &&  $<$ 23.9 & S1 & (D2, T3) &  \\
      12 &    &    &    &  16:08:03  & & 2.24 && S1 &  & \\
      13 &    &    &    &  18:32:55  & & 2.41 && T1  &  & \\
      14 &    &    &    &  18:43:22  & & 0.174 && T2  &  & \\
      15 &    &    &    &  18:55:05  & & 0.195 && T3  &  & \\
      16 &    &    &    &  23:15:33  & & 4.34 && S1 & (D2)	&  \\
      17 &    &    &  22  &  02:09:52  & & 2.91 && S1 &  & \\
      18 &    &    &    &  04:26:17  & & 2.27 && D1  &  & \\
      19 &    &    &    &  04:35:11  & & 0.148 && D2  &  & \\
      20 &    &    &    &  07:10:07  & & 2.58 && S1 & (D2)	&  \\
      21 &    &    &    &  09:23:54  & & 2.23 && S1 &  & \\
      22 &    &    &    &  11:37:24  & & 2.23 && T1  &  & \\
      23 &    &    &    &  11:51:07  & & 0.229 && T2  &  & \\
      24 &    &    &    &  12:04:08  & & 0.217 && T3  &  & \\
      25 &  694  &    &  23  &  12:32:29  &  1.83 $<$  &&  $<$ 24.5 & S1 &  & \\
      26 &    &    &    &  15:05:56  & & 2.56 && S1 &  & \\
      27 &    &    &    &  17:25:09  & & 2.32 && S1 & (D2)	&  \\
      28 &    &    &    &  20:00:08  & & 2.58 && S1 &  & \\
      29 &    &    &    &  22:34:36  & & 2.57 && D1  &  & \\
     30 &    &    &    &  22:45:51  & & 0.187 && D2  &  & \\
     31 &    &    &  24  &  02:10:45  & & 3.41 && D1  &  & \\
     32 &    &    &    &  02:26:15  & & 0.259 && D2  &  & \\
     33 &    &    &    &  06:03:06  & & 3.61 && S1 &  & \\
     34 &    &    &    &  08:50:34  & & 2.79 && T1  &  & \\
     35 &    &    &    &  09:05:55  & & 0.256 && T2  &  & \\
     36 &    &    &    &  09:16:31  & & 0.177 && T3  &  & \\
     37 &    &    &    &  13:36:43  & & 4.34 && S1 &  & \\
     38 &    &    &    &  16:37:55  & & 3.02 && S1 & (D1, T1)	&  \\
$^{a}$39 &  695  &    &  25  &  17:28:20  &  0.036 $<$  &&  $<$ 24.8 & D1  & (T2)	&  \\
     40 &    &    &    &  17:41:31  & & 0.220 && D2  & (T3)	&  \\
     41 &    &    &    &  21:34:54  & & 3.89 && S1 &  & \\
     42 &    &    &  26  &  00:28:00  & & 2.89 && D1  &  & \\
     43 &    &    &    &  00:46:20  & & 0.306 && D2  &  & \\
     44 &    &    &    &  04:28:35  & & 3.70 && D1  &  & \\
     45 &    &    &    &  04:44:35  & & 0.267 && D2  &  & \\
     46 &    &    &    &  08:51:35  & & 4.12 && S1 &  & \\
     47 &    &    &    &  11:41:43  & & 2.84 && S1 &  & \\
     48 &  708  &  Oct  &  21  &  11:51:35  &  1.82 $<$  &&  $<$ 600 & S1 & (D1)	&  \\
     49 &    &    &    &  14:12:10  & & 2.34 && S1 & (D1)	&  \\
     50 &    &    &    &  16:30:24  & & 2.30 && T1  &  & \\
     51 &    &    &    &  16:39:49  & & 0.157 && T2  &  & \\
     52 &    &    &    &  16:51:05  & & 0.188 && T3  &  & \\
     53 &    &    &    &  19:59:49  & & 3.15 && D1  &  & \\
     54 &    &    &    &  20:18:52  & & 0.317 && D2  &  & \\
     55 &    &    &  22  &  00:28:46  & & 4.17 && S1 &  & \\
     56 &  710  &    &  25  &  20:58:42  &  1.65 $<$  &&  $<$ 92.5 & S1 &  & \\
     57 &    &    &    &  23:48:23  & & 2.83 && D1  &  & \\
     58 &    &    &  26  &  00:01:37  & & 0.221 && D2  &  & \\
     59 &    &    &    &  03:11:08  & & 3.16 && S1 & (D1)	&  \\
     60 &    &    &    &  05:38:40  & & 2.46 && S1 & (D2)	&  \\
     61 &    &    &    &  07:41:16  & & 2.04 && D1  &  & \\
     62 &    &    &    &  07:53:28  & & 0.203 && D2  &  & \\
     63 &    &    &    &  10:58:02  & & 3.08 && D1  &  & \\
     64 &    &    &    &  11:13:05  & & 0.251 && D2  &  & \\
     65 &  719  &  Nov  &  12  &  08:35:44  &  0.191 $<$  &&  $<$ 405 & S1 & (D2, T3)	&  \\
     66 &    &    &    &  11:42:47  & & 3.12 && S1 &  & \\
     67 &    &    &    &  15:23:00  & & 3.67 && S1 &  & \\
     68 &    &    &    &  18:31:01  & & 3.13 && S1 &  & \\
     69 &    &    &    &  21:17:46  & & 2.78 && D1  &  & \\
     70 &    &    &    &  21:26:09  & & 0.140 && D2  &  & \\
     71 &    &    &  13  &  01:04:58  & & 3.65 && S1 &  & \\
     72 &    &    &    &  03:39:57  & & 2.58 && T1  &  & \\
     73 &    &    &    &  03:51:47  & & 0.197 && T2  &  & \\
     74 &    &    &    &  04:03:36  & & 0.197 && T3  &  & \\
     75 &    &    &    &  07:27:43  & & 3.40 && D1  &  & \\
     76 &    &    &    &  07:39:37  & & 0.198 && D2  &  & \\

\hline
\noalign {\smallskip}
\multicolumn{9}{l}{$^{a}$ detected by RGS only, before EPIC cameras were turned on.}\\
\end{tabular}
\end{center}
\label{atab:times}
\end{table*}

\end{appendix}

\end{document}